# Red material on the large moons of Uranus: Dust from the irregular satellites?

Richard J. Cartwright[1], Joshua P. Emery[2], Noemi Pinilla-Alonso[3], Michael P. Lucas[2], Andy S. Rivkin[4], and David E. Trilling[5].

*[1]Carl Sagan Center, SETI Institute; [2]University of Tennessee; [3]University of Central Florida; [4]John Hopkins University Applied Physics Laboratory; [5]Northern Arizona University.*

**Abstract**

The large and tidally-locked "classical" moons of Uranus display longitudinal and planetocentric trends in their surface compositions. Spectrally red material has been detected primarily on the leading hemispheres of the outer moons, Titania and Oberon. Furthermore, detected $H_2O$ ice bands are stronger on the leading hemispheres of the classical satellites, and the leading/trailing asymmetry in $H_2O$ ice band strengths decreases with distance from Uranus. We hypothesize that the observed distribution of red material and trends in $H_2O$ ice band strengths results from infalling dust from Uranus' irregular satellites. These dust particles migrate inward on slowly decaying orbits, eventually reaching the classical satellite zone, where they collide primarily with the outer moons. The latitudinal distribution of dust swept up by these moons should be fairly even across their southern and northern hemispheres. However, red material has only been detected over the southern hemispheres of these moons, during the Voyager 2 flyby of the Uranian system (subsolar latitude ~81ºS). Consequently, to test whether irregular satellite dust impacts drive the observed enhancement in reddening, we have gathered new ground-based data of the now observable northern hemispheres of these moons (sub-observer latitudes ~17 – 35ºN). Our results and analyses indicate that longitudinal and planetocentric trends in reddening and $H_2O$ ice band strengths are broadly consistent across both southern and northern latitudes of these moons, thereby supporting our hypothesis. Utilizing a suite of numerical best fit models, we investigate the composition of the reddening agent detected on these moons, finding that both complex organics and amorphous pyroxene match the spectral slopes of our data. We also present spectra that span L/L' bands (~2.9 – 4.1 µm), a previously unexplored wavelength range in terms of spectroscopy for the Uranian moons, and we compare the shape and albedo of the spectral continua in these L/L' band data to other icy moons in the Jovian and Saturnian systems. Additionally, we discuss possible localized enhancement of reddening on Titania, subtle differences in $H_2O$ ice band strengths between the southern and northern hemispheres of the classical satellites, the distribution of constituents on Miranda, and the possible presence of $NH_3$-hydrates on these moons. In closing, we explore potential directions for future observational and numerical modeling work in the Uranian system.

## 1. Introduction

Ground-based observations of the large and tidally-locked "classical" Uranian satellites Miranda, Ariel, Umbriel, Titania, and Oberon (Table 1) have revealed that the surfaces of these moons are dominated by $H_2O$ ice, intimately mixed with a low albedo and spectrally neutral constituent (*e.g.*, Cruikshank, 1980; Cruikshank and Brown, 1981; Soifer et al., 1981; Brown and Cruikshank, 1983; Brown and Clark, 1984). Laboratory experiments suggest that this low albedo material is carbonaceous in composition, spectrally similar to charcoal (*e.g.*, Clark and Lucey, 1984). More recent ground-based observations have detected "pure" $CO_2$ ice (*i.e.*, molecularly bound to other $CO_2$ molecules), segregated from other constituents (*i.e.*, surrounded primarily by other $CO_2$ ice grains), primarily on the trailing hemispheres of the moons closest to Uranus (Grundy et al., 2003, 2006; Cartwright et al., 2015). The combination of $H_2O$ ice-rich regions intimately mixed with dark material, along with regions of pure and segregated $CO_2$ ice, makes this group of moons fairly distinct from other icy objects in the outer Solar System.

The Imaging Science System (ISS, ~0.4 – 0.6 µm) onboard Voyager 2 collected spatially-resolved images of these satellites during its brief encounter with the Uranian system (*e.g.*, Smith et al., 1986; Stone and Miller, 1986) (Figure 1). Analysis of those ISS images demonstrated that the leading hemispheres (1 – 180º longitude) of the Uranian moons are spectrally redder than their trailing hemispheres (181 – 360º longitude) (Bell and McCord, 1991; Buratti and Mosher, 1991; Helfenstein et al., 1991). Furthermore, the degree of reddening increases with distance from Uranus, with minimal reddening of the innermost moon Miranda, moderate reddening of the middle moons, Ariel and Umbriel, and substantial reddening of the outer moons, Titania and Oberon (Buratti and Mosher, 1991).

The observed longitudinal and planetocentric trends in reddening on these moons likely result from system-wide surface modification processes. There are three broad types of system-wide mechanisms that operate in giant planet systems: charged particle bombardment, *inter*planetary (heliocentric) dust impacts, and *intra*planetary (planetocentric) dust impacts. Although all three of these processes can redden surfaces (*e.g.*, Bennett et al., 2013, and references therein), the observed longitudinal and planetocentric trends for each process should be distinct. The faster rotation rate of Uranus' magnetic field (~17.2 hours) compared to the orbits of the classical moons (~1.5 to 13.5 days) and greater magnetic field densities closer to Uranus (Ness et al., 1986) should increase charged particle bombardment into (and enhance the reddening of) the trailing hemispheres of the moons closest to Uranus. Due to gravitational focusing by Uranus, and the faster orbital velocities of the inner moons, heliocentric dust should preferentially impact (and enhance the reddening of) the leading hemispheres of the moons closest to Uranus (Zahnle et al., 2003). Planetocentric dust particles infalling toward Uranus on decaying orbits should preferentially impact (and enhance the reddening of) the leading hemispheres of the outer classical moons (Tamayo et al., 2013a). Furthermore, the nine known irregular satellites of Uranus are all redder than the five classical moons (*e.g.*, Gladman et al., 1998; Grav et al., 2003, 2004; Maris et al., 2002, 2007). Thus, dust from the Uranian irregular satellites represents the most plausible source of red material that can generate planetocentric and longitudinal trends consistent with the observed trends in reddening on the classical moons.

Measurements of the 1.52-µm and 2.02-µm $H_2O$ ice bands present in Uranian satellite spectra indicate that $H_2O$ ice bands are stronger on the leading hemispheres of these moons compared to their trailing hemispheres (Grundy et al., 2006; Cartwright et al., 2015). Furthermore, the longitudinal asymmetry in $H_2O$ band strengths decreases with distance from Uranus (Grundy et al., 2006; Cartwright et al., 2015). This longitudinal trend is consistent with greater accumulation of spectrally red dust, which could help mask the spectral signature of $H_2O$ ice on the leading hemispheres of these satellites, in particular on the outer moons. Thus, based on the longitudinal and planetocentric trends in reddening and $H_2O$ band strengths, we hypothesize that the red material observed on the classical moons originated on the surfaces of retrograde irregular satellites. The degree of reddening with this mechanism should be similar over southern and northern latitudes of the classical moons, without substantial latitudinal preference.

To test this hypothesis, we have gathered new spectral data of the now observable northern hemispheres of the classical Uranian satellites (sub-observer latitudes ~17 – 35°N). The Voyager 2 encounter with Uranus occurred near southern summer solstice, when the subsolar point was ~81°S, and ISS almost exclusively imaged the southern hemispheres of the Uranian satellites (Figures 1 and 2). Additionally, previous analyses of $H_2O$ ice band strengths were made using ground-based, near-infrared (NIR) spectra collected over the southern hemispheres of these moons (~6 – 30°S) (Grundy et al., 2006). Therefore, measurement of spectral slopes and $H_2O$ ice band strengths in our northern hemisphere data, and comparison to these same measurements in previously collected southern hemisphere data, will enable us to test whether the red material originated on retrograde irregular satellites. We precede the presentation of our work with some necessary background on heliocentric and planetocentric dust.

## 2. Background

Impact events on irregular satellites eject dust particles off their surfaces and into orbit around their giant planet primaries. These ejected dust particles initially have the same orbital elements (semi-major axes, eccentricities, inclinations, etc) as the irregular moon they originated on. However, over ~5 to 500 Myr timescales (for 10 and 1000 µm grain sizes, respectively), the semi-major axes of these dust particles decay due to Poynting-Robertson drag (*i.e.*, loss of forward momentum due to isotropic re-radiation of solar flux) (Burns et al., 1979), and the dust slowly migrates inward. For low obliquity planets like Jupiter and Saturn, almost all of this planetocentric dust collides with the outermost classical satellite in each system (Callisto and Iapetus, respectively) (Bottke et al., 2013).

For high obliquity ($\gtrsim 70^{\circ}$) planetary systems like Uranus, the inward migration of planetocentric dust is likely more complicated. In the Uranian system, infalling planetocentric dust particles encounter an unstable "chaotic" region as their semi-major axes decay (*e.g.*, Tremaine et al., 2009; Tamayo et al., 2013b), causing their inclinations and eccentricities to experience large amplitude oscillations over short timescales (~$10^4$ yrs). Numerical simulations by Tamayo et al. (2013a) indicate that the fates of the dust particles depend on their initial orbits (*i.e.*, the orbits of the irregular satellite from which they originate) and the size of the dust grains. Large dust grains (> 50 µm) with low initial eccentricity ($e_o \lesssim 0.20$) impact Ariel, Umbriel, Titania, and Oberon with fairly even probabilities. In all other cases (small grains and all dust grains with $e_o \gtrsim 0.20$), the collision probabilities decrease with decreasing planetocentric distance (*i.e.*, are higher at Oberon and decrease systematically to Ariel). This latter situation is most applicable to the Uranus system. Seven of the nine known irregular moons have orbital eccentricities > 0.20. This includes the largest irregular moon, Sycorax, with an eccentricity of 0.52. Furthermore, analysis of Cassini imaging of the Phoebe dust ring demonstrates that the size distribution of that dust is very steep, strongly dominated by small grains (< 20 µm) (Hamilton et al., 2015; Tamayo et al., 2017). Assuming dust production mechanism at Saturn and Uranus are similar, then the dust colliding with the classical moons could display a similar dominance by small grains. Taking these considerations into account, the numerical simulations by Tamayo et al. (2013a) predict that the total amount of irregular satellite dust deposition scales with planetocentric distance, with Oberon receiving the most dust (over a wide range of modeled grain sizes), except for large dust grains with low initial eccentricities, which collide with each of these moons with fairly even probabilities.

A leading/trailing dichotomy in dust collisions is also expected. Eight of the nine known Uranian irregular satellites have retrograde orbits. The Tamayo et al. (2013a) simulations demonstrate that dust particles from retrograde irregular satellites collide with the leading hemispheres of the classical moons at a higher frequency (factor of ~2 – 3) than their trailing hemispheres. This dust mantle should display a fairly homogenous distribution across both the southern and northern latitudes of these moons. The net result of the Tamayo et al. (2013a) models is that dust from retrograde irregular satellites should collide primarily with the leading hemispheres of Oberon, Titania, Umbriel, and Ariel in order of decreasing probability, respectively, with negligible amounts of dust surviving to collide with the innermost moon Miranda.

The Uranian moons have likely accumulated substantial amounts of irregular satellite dust over the age of the Solar System. Dynamical simulations indicate that when the irregular satellite swarms of the giant planets were initially captured and/or experienced subsequent orbital perturbations, these objects frequently collided, generating large amounts of dust (Bottke et al., 2010). Much of this primordial dust would have collided with the classical Uranian moons early on in Solar System history. However, recent impacts and collisions between irregular satellites can also generate substantial amounts of dust, as evidenced by the torus of material detected around the Saturnian moon Phoebe (Verbiscer et al., 2009; Tamayo et al., 2014). Dust from Phoebe, and the other Saturnian irregular satellites, is swept up primarily by the leading hemisphere of the outer classical moon, Iapetus, with a small amount reaching the orbits of Titan and Hyperion (*e.g.*, Tamayo et al., 2011).

Unlike planetocentric dust, heliocentric dust particles are not gravitationally bound to Uranus on slowly decaying orbits. Instead, heliocentric micrometeorites enter the Uranian system as high velocity "bullets," heading for collisions primarily with the inner moons due to gravitational focusing by Uranus (Tamayo et al., 2013a). Furthermore, the faster orbital velocities of the inner moons should increase the frequency of heliocentric dust collisions with their leading hemispheres compared to their trailing hemispheres, unlike the slower outer moons, which should exhibit smaller leading/trailing asymmetries in heliocentric dust impacts. The larger leading/trailing asymmetries in $H_2O$ ice band strengths for the moons closer to Uranus are therefore consistent with enhanced regolith overturn by heliocentric dust grains, assuming the near-surfaces of these moons are $H_2O$ ice-rich (*e.g.*, Bennett et al., 2013; Clark et al., 2013 and references therein). The likely lower velocity of planetocentric dust impacts compared to heliocentric dust impacts (few km $s^{-1}$ and ~10 – 30 km $s^{-1}$, respectively) should reduce regolith overturn and promote the accumulation of a layer of planetocentric dust, primarily on the outer moons. Therefore, the combination of fewer heliocentric dust impacts and more planetocentric dust impacts on the outer moons should reduce leading/trailing $H_2O$ band asymmetries compared to the inner moons.

## 3. Observations and Data Reduction

Uranian satellite observations were made using the NIR SpeX spectrograph/imager at †NASA's Infrared Telescope Facility (IRTF) on Mauna Kea (Rayner et al., 1998, 2003). SpeX spectra were collected by four different teams (summarized in Tables 2 and 3 and displayed in Figure 2). In total, we have analyzed 90 SpeX spectra of the classical Uranian moons (28 collected over their southern hemispheres and 62 over their northern hemispheres). Observing details for each reduced spectrum, including mid-observation satellite latitudes and longitudes, are summarized in Table 3. Observations made by Grundy were presented in Grundy et al. (2003, 2006), observations made by Gourgeot were presented in Gourgeot et al. (2014), and spectra collected by Rivkin and Cartwright (2012 – 2013) were presented in Cartwright et al. (2015). We refer the reader to these four papers for information regarding each team's data reduction routines. Here, we present 41 new SpeX spectra of the Uranian satellites gathered in PRISM (low resolution, single order) and SXD (moderate resolution, dispersed across multiple orders) modes, with wavelength ranges spanning ~0.7 to 2.5 μm. We also present six new SpeX spectra gathered in LXD "short" mode (moderate resolution, dispersed across multiple orders), which covers a wavelength range of ~1.7 to 4.2 μm. The spectral resolution and wavelength range of each mode is summarized in Table 2.

All data were collected utilizing the 'AB' nodding mode of SpeX, where the target spectrum is measured in two different positions on the array ('A' and 'B'), separated by 7.5 arcsec along the 15 arcsec slit. First order sky emission correction was performed by subtracting the 'A' position frame from the 'B' position frame. Each team limited individual frame exposure times to a maximum of 120 seconds for SXD and PRISM mode spectra to minimize the impact of atmospheric variability on gathered data.

Individual frame exposure times for LXD mode spectra were limited to 20 seconds to avoid saturation at longer wavelengths. Flat field and wavelength calibration frames were generated using a quartz lamp and argon lamp, respectively, with SpeX's internal integrating sphere. Because argon emission lines are quite weak at wavelengths > 2.5 µm, night sky emission lines are also used to calibrate the longer wavelength orders of LXD mode spectra.

The 41 new spectra were extracted, calibrated, and background subtracted using a combination of custom codes and the Spextool data reduction package (Cushing et al., 2004; Vacca et al., 2003). To boost signal-to-noise (S/N), all frames for each object observed on a given night, collected using the same mode of SpeX, were co-added during reduction. Selected analog stars (Table 4) were observed at multiple times throughout the night, over a range of airmasses. Satellite spectra were divided by solar analog spectra from the same night, observed at similar airmasses (within ± 0.1 airmasses in most cases), to remove the solar spectrum and correct for telluric absorptions. These reduced spectra were combined using custom programs and the Spextool program suite (Cushing et al., 2004).

To remove residual telluric contamination, we performed subpixel shifting of star and object spectra using custom codes. Additional correction routines were applied to some of the reduced spectra, including interpolation of star airmasses to more precisely match object airmasses and division by an appropriately scaled atmospheric transmission spectrum collected over Mauna Kea. To correct for scattered light from Uranus between ~0.7 and 1.5 µm in some spectra of Ariel and Umbriel, and all Miranda spectra, we used a scaled SpeX spectrum of Uranus gathered each night we observed these moons.

We scaled the PRISM and SXD spectra of these moons to their geometric albedos at ~0.96 µm (Karkoschka, 2001). Longer wavelength LXD spectra were then scaled to PRISM and SXD spectra using the wavelength regions where these spectral datasets overlap (~1.7 – 2.55 µm). Because the classical Uranian satellites do not display systematic hemispherical variations in their albedos (Veverka et al., 1991), we utilized the same geometric albedo for their leading and trailing hemispheres. We present 'grand average' PRISM mode spectra for each moon, along with the six longer wavelength LXD mode spectra we have gathered for Ariel, Titania, and Oberon, in Figure 3. We compare these LXD spectra to broadband geometric albedos calculated using images gathered by the Infrared Array Camera (IRAC, 4 channels spanning ~3.1 to 9.5 µm) (Fazio et al., 2004) onboard the Spitzer Space Telescope (Werner et al., 2004) in the Analysis section of this paper (IRAC geometric albedos were originally presented in Cartwright et al., 2015).

## 4. Methods

### *4.1. Leading and trailing hemisphere reflectance ratios*

With these new NIR spectra, we quantified the degree of reddening by hemisphere on each moon. Because most of the dust from the irregular satellites, over a wide range of grain sizes, is predicted to collide with the outer moons, we interpret the observed degree of reddening as an assessment of the total amount of dust accumulated on each moon. Although the SXD and PRISM modes of SpeX cover similar wavelength ranges (~0.7 – 2.5 µm), the cross-dispersed SXD mode spectra have lower S/N at wavelengths < 1.2 µm compared to PRISM mode spectra, and we therefore conducted our slope analyses using only PRISM spectra.

The wavelength range where reddening is apparent on these moons spans ~0.7 to 1.3 µm (see section *5.1* for a detailed description of the analyzed spectra). We utilized reflectance ratios to quantify the degree of reddening over this wavelength range. We selected 0.724 µm as the short wavelength end for our reflectance ratio measurements, ensuring high S/N and avoiding any possible 'edge effects' at the short wavelength end of the spectra (~0.67 µm). The analyzed PRISM spectra display an increase in reflectance with wavelength (*i.e.*, spectral reddening) up to ~1.25 µm, and then at longer wavelengths, the spectral continua display neutral slopes until absorption by the 1.52-µm $H_2O$ ice band starts near 1.4 µm. Consequently, we set the long wavelength end of our reflectance ratio measurements to 1.260 µm, capturing the entire wavelength range where reddening is apparent (~0.72 – 1.25 µm). We averaged an equal number of reflectance data points together at either end of the selected wavelength range: between 0.724 and 0.740 µm ($R_{0.72\ \mu m}$) and 1.230 and 1.260 µm ($R_{1.26\ \mu m}$).

To estimate the uncertainties for these mean reflectances, we first calculated the mean error by adding the $1\sigma$ uncertainties for each spectrum in quadrature and dividing by the sample size ($n$). In order to capture the point-to-point variation of the spectra in our uncertainty estimates, we calculated the standard deviation of the mean ($\sigma_{\bar{x}} = \sigma/\mathrm{sqrt}(n)$) for both ends of the selected wavelength range. Next, we added the mean error and $\sigma_{\bar{x}}$ in quadrature ($\Delta R_{0.72\ \mu m}$ and $\Delta R_{1.26\ \mu m}$) for each spectrum, thereby calculating the total error for both ends of the selected wavelength range. Finally, we divided the mean $R_{1.26\ \mu m}$ by the mean $R_{0.72\ \mu m}$ ($R_{1.26\ \mu m}/ R_{0.72\ \mu m}$) for each PRISM spectrum and propagated the errors for these mean reflectances (*e.g.*, Taylor, 1982). The wavelength range of the reflectance ratio measurements is illustrated in Figure 4. Because of substantial Uranian scattered light contamination over this wavelength range in the Miranda spectra, we did not calculate reflectance ratios for this innermost classical moon.

### *4.2. Leading and trailing hemisphere $H_2O$ band area and depth measurements*

The spectral signature of $H_2O$ ice includes multiple overlapping combination and overtone band complexes in the NIR region covered by SpeX spectra, centered near 0.81, 0.90, 1.04, 1.25, 1.52, and 2.02 µm, with an additional absorption feature present in crystalline ice-rich surfaces centered near 1.65 µm (*e.g.*, Grundy and Schmitt, 1998; Mastrapa et al., 2008). Spectral analysis of $H_2O$ ice intimately mixed with dark material (both silicate and carbonaceous species) demonstrates that the strengths of these $H_2O$ ice bands are greatly reduced by relatively minor amounts of dark contaminants (*e.g.*, Clark and Lucey, 1984). Spectra collected over the low albedo, ice-rich surfaces of the Uranian satellites provide clear evidence of the 1.52-µm band, including the crystalline $H_2O$ feature centered near 1.65 µm, and the 2.02-µm band, with no evidence for the weaker $H_2O$ bands between ~0.8 and 1.25 μm (Figure 4). We therefore focused our analysis on the 1.52-μm and 2.02-μm band complexes.

Following Cartwright et al. (2015), we measured the areas and depths of the 1.52-μm and 2.02-μm bands using a modified version of the Spectral Analysis Routine for Asteroids program (SARA, Lindsay et al., 2015). First, we defined the continuum on either side of both bands (Table 5), taking particular care on the short wavelength end of both $H_2O$ band complexes to avoid any residual telluric contributions. The long wavelength end of the 2.02-μm band was defined using a third order polynomial to fit the 'arched' $H_2O$ ice continuum between 2.215 and 2.230 μm. We then connected the short and long wavelength ends of the bands' continua with a line, and divided both bands by their associated continuum (see example in Figure 4), thereby normalizing these $H_2O$ bands prior to taking measurements. Using the trapezoidal rule, we calculated the integrated 1.52-μm and 2.02-μm $H_2O$ band areas for each spectrum. We estimated the $1\sigma$ errors for our band area measurements by running Monte Carlo simulations, resampling from within a Gaussian distribution, represented by the $1\sigma$ error bars for each spectral channel. This process was iterated 20,000 times. For more details on this technique, see Lindsay et al. (2015).

To measure the depths of these two bands, we calculated the mean, continuum-divided reflectance ($R_{band}/R_{continuum}$) between 1.515 and 1.525 μm and 2.015 and 2.025 μm (for the 1.52-μm and 2.02-μm bands, respectively), and subtracted these mean values from 1.0 (wavelength ranges for these band center measurements are shown in Figure 4). To estimate uncertainties for these depth measurements, we calculated a mean error by adding the $1\sigma$ errors of the individual spectra in quadrature and dividing by the sample size ($n$), and then we calculated $\sigma_{\bar{x}}$ to estimate point-to-point variation for the band center data points. Finally, we added the mean error and $\sigma_{\bar{x}}$ in quadrature for each band center measurement.

## 5. Results

### *5.1. Detected absorption features and spectral slopes*

We detected the 1.52-μm and 2.02-μm $H_2O$ ice band complexes in all of the 41 new SXD and PRISM spectra presented here (see Figures 3, 4, and Appendix A). The absence of weak $H_2O$ bands between ~0.8 and 1.25 μm is consistent with previous analyses of NIR reflectance spectra of these moons (*e.g.*, Grundy et al., 2006; Cartwright et al., 2015). There is strong evidence for the three $CO_2$ ice combination and overtone absorption bands centered near 1.966, 2.012, and 2.070 μm in the moderate resolution SXD spectra collected over the trailing hemispheres of these moons (in particular for Ariel), consistent with prior work (Grundy et al., 2003, 2006; Cartwright et al., 2015). Updated analyses of these $CO_2$ ice absorption bands will be reported in future work. We note subtle, but apparent, spectral reddening between ~0.7 and 1.3 μm for Ariel, Umbriel, Titania, and Oberon. For all five moons, we see evidence of spectral 'blueing' (*i.e.*, reduction in reflectance with wavelength) between ~1.4 and 2.5 µm, consistent with the presence of $H_2O$ ice (*e.g.*, Clark et al., 2013, and references therein).

Some of the spectra analyzed in this study hint at the presence of an absorption feature centered near 2.2 µm (Figure 5). The spectral characteristics of this feature are qualitatively consistent with the position and shape of the 2.2 µm feature detected on Charon, which has been attributed to $NH_3$-hydrates (*e.g.*, Buie and Grundy, 2000; Brown and Calvin, 2000; Dumas et al., 2001; Cook et al., 2007; Grundy et al., 2016, Cook et al., 2017; Dalle Ore et al., 2018). A previous study has attributed a similar feature centered near 2.2 µm in spectra of Miranda to $NH_3$-hydrates as well (Bauer et al., 2002). We consider the implications of the possible detection of $NH_3$-hydrates on these moons in section *7.5*.

The native S/N of the LXD spectra in L/L' bands are quite low (S/N of ~1.5 to 3 between ~2.9 and 4.1 µm). Thus, we binned these spectra by a factor of 50, sufficiently increasing their S/N (~10 – 20) to characterize the shapes and relative reflectances of their continua over this wavelength range (native and binned LXD spectra shown in Figure 3). We detect clear evidence for the 3-µm $H_2O$ ice band (Mastrapa et al., 2009) in all six LXD spectra. Additionally, the spectra of Titania and the trailing hemisphere of Oberon hint at the presence of a subtle Fresnel peak centered near 3.1 µm, consistent with the presence of crystalline $H_2O$ ice (Mastrapa et al., 2009). The LXD spectrum of Oberon's leading hemisphere also hints at the presence of an $H_2O$ ice Fresnel peak, but this feature is centered closer to 3.05 µm, which is more consistent with amorphous $H_2O$ ice (Mastrapa et al., 2009). Evidence for a Fresnel peak in the LXD spectra of Ariel is less clear, albeit, the S/N of the Ariel spectra are lower than the S/N of the Titania and Oberon spectra. All six LXD spectra display a steep increase in reflectance on the long wavelength end of the 3-µm $H_2O$ ice band, peaking between ~3.5 to 3.6 µm. The reflectance of the spectra then steadily decreases toward the long wavelength edge of each spectrum (~4.1 µm), presumably due to absorption by the $H_2O$ ice combination band centered near 4.5 µm. We compare these LXD spectra to channel 1 IRAC geometric albedos (presented in Cartwright et al., 2015) in section *6.5*.

### *5.2. Reflectance ratios*

The $R_{1.26\,\mu m}/\,R_{0.72\,\mu m}$ results (Table 6) indicate an increase in reddening with distance from Uranus, with more pronounced reddening on the outer moons, Titania and Oberon. The leading and trailing hemispheres of Ariel and Umbriel do not display discernable asymmetries in their reflectance ratios, with similar levels of reddening on both their leading and trailing hemispheres. In contrast, the reflectance ratios for the outer moons Titania and Oberon show that their leading hemispheres are redder than their trailing hemispheres (> $2\sigma$ for most of the reflectance ratios). We present further analysis of our reflectance ratio results in section *6.1*.

### *5.3. $H_2O$ band areas and depths*

Our $H_2O$ ice band parameter measurements (with $1\sigma$ uncertainties) are summarized in Table 7. Consistent with previous analyses of $H_2O$ ice bands on these moons (Grundy et al., 2006; Cartwright et al., 2015), our band parameter results indicate that the detected $H_2O$ bands are mostly stronger (*i.e.*, larger band areas and depths) on the leading hemispheres of these moons, compared to their trailing hemispheres. Our measurements indicate that the strongest $H_2O$ bands are in spectra of the leading hemisphere of Ariel, and the weakest bands are in spectra of Umbriel. When considering only the trailing hemispheres of the Uranian satellites, the Titania spectra display the strongest $H_2O$ bands, with the weakest bands in the Umbriel spectra. Furthermore, the asymmetry in $H_2O$ band strengths between the leading and trailing hemispheres of these moons decreases with distance from Uranus. Unlike the planetocentric and longitudinal trends in $H_2O$ band strengths for Ariel, Umbriel, Titania, and Oberon, the measured $H_2O$ ice bands appear to be slightly stronger on the trailing hemisphere of Miranda. We explore our $H_2O$ ice band measurements further in section *6.3*.

## 6. Analysis

In order to constrain the distribution of red material on the classical Uranian satellites, we conducted statistical analyses of our reflectance ratio and $H_2O$ band area results. We focused these analyses on the distribution of these constituents as a function of satellite longitude and as a function of planetocentric distance.

### *6.1. Longitudinal and planetocentric trends in reflectance ratios*

To test whether the leading and trailing hemispheres of these moons display statistically significant differences in spectral reddening, we compared our reflectance ratio results (weighted by their uncertainties) to two different models: mean and sinusoidal (solid purple and black dashed lines in Figure 6). The mean model is an average reflectance ratio for each moon, representing a surface without significant variations in spectral reddening. The sinusoidal model (utilizing three coefficients: amplitude, phase shift, and vertical offset) represents longitudinal asymmetries in the distribution of spectrally red material on these moons. To compare these models, we used an $F$-test (*e.g.*, Speigel et al., 1992), with the null hypothesis that there is no statistically significant difference between the fits provided by the mean and sinusoidal models.

The $F$-test analyses demonstrate that there is no significant difference between these two models for the inner moons, Ariel and Umbriel, supporting the null hypothesis (Table 8). For Titania and Oberon, on the other hand, the $F$-test results indicate that the sinusoidal model is a significantly better fit ($> 2\sigma$) for the reflectance ratio measurements, and we reject the null hypothesis for these two satellites (Table 8).

To explicitly look for longitudinal-scale variations of red material on these moons, we also calculated mean reflectance ratios for the leading and trailing hemisphere of each moon, reported in Table 9 and Figure 7a with $2\sigma$ uncertainties. We also report the degree of reddening on these moons in terms of spectral gradient (per 0.1 μm) to facilitate comparison with other red icy objects elsewhere in the outer Solar System (Table 9) (*e.g.*, Jewitt et al., 2002; Doressoundiram et al., 2008). To highlight the trends in reddening on these satellites, we divided the mean reflectance ratios of their leading hemispheres by the mean reflectance ratios of their trailing hemispheres (Figure 7b).

Similar to our *F*-test analyses of the individual reflectance ratios, the mean reflectance ratios reveal that there are only minor variations in reddening between the leading and trailing hemispheres of the inner moons, Ariel and Umbriel, but clear differences ($> 2\sigma$) between the leading and trailing hemispheres of the outer moons, Titania and Oberon. When comparing the mean reflectance ratios for the trailing hemispheres of Umbriel, Titania, and Oberon, we find that these ratios are all statistically similar ($< 2\sigma$ difference), whereas the trailing hemisphere of Ariel is clearly less red than the other moons (Figure 7a). Thus, the increased reddening of the outer moons compared to the inner moons is most apparent when comparing their leading hemispheres. Consequently, our reflectance ratio analyses support the hypothesis that spectrally red dust from Uranus' irregular satellites preferentially accumulates on the leading hemispheres of the outer moons.

### *6.2. Latitudinal trends in spectral reddening*

Along with constraining the longitudinal distribution of red material, we investigated the degree of reddening over the southern and northern latitudes of these moons. To do this, we compared our mean reflectance ratios (subsolar latitudes ~25 – 35°N) to the Voyager 2-era color ratios (subsolar latitude ~81°S), calculated using the violet and green filters of ISS (spanning ~0.4 to 0.6 µm). In Table 10 we compare the mean leading/trailing PRISM reflectance ratios (shown in Figure 7b) to the leading/trailing ISS color ratios reported in Buratti and Mosher (1991).

Overall, the two datasets are broadly consistent, showing that the outer moons are redder than the inner moons (mean PRISM reflectance ratios are shown in Table 9, average ISS Gr/Vi color ratios are shown in column 3 of Table 10). Furthermore, both datasets show that the inner moons display only minor differences in reddening between their leading and trailing hemispheres, whereas the leading hemisphere of Oberon is clearly redder than its trailing hemisphere. The leading hemisphere of Titania is also clearly redder than its trailing hemisphere over its northern latitudes, but the leading/trailing asymmetry in reddening over Titania's southern latitudes is subtler. In general, reddening is somewhat more pronounced in the visible (VIS) wavelength ISS dataset compared to the NIR wavelength PRISM dataset, which could possibly result from a change in spectral slope between these two wavelength regions (*i.e.*, the red material is redder in the VIS than the NIR).

### *6.3. Longitudinal and planetocentric trends in $H_2O$ band areas*

We investigated the areas of the 1.52-µm and 2.02-µm $H_2O$ ice bands as a function of satellite longitude for all five satellites (Figure 8). We used an *F*-test to compare mean and sinusoidal models fit to the $H_2O$ band area data points for each moon (solid purple and black dashed lines in Figure 8). The results of this *F*-test analysis (Table 8) indicate that there is a significant difference ($> 2\sigma$) for the 1.52-µm and 2.02-µm $H_2O$ band areas between the leading and trailing hemispheres of Ariel, Umbriel, Titania, and Oberon. Unlike the other moons, Miranda appears to display a significant difference ($> 2\sigma$) in $H_2O$ band strength between its anti-Uranus (135 – 225° longitude) and Uranus-facing (315 – 45° longitude) quadrants for the 1.52-µm $H_2O$ ice band (but not the 2.02-µm band). Curiously, the one available data point near the predicted peak at 180° longitude (mid-observation longitude 165.0º) falls well below the model fit. We intend to follow up with additional observations of Miranda to gain a better understanding of the distribution of $H_2O$ band strengths for this moon.

Using the individual $H_2O$ band area measurements (Table 7), we calculated mean $H_2O$ band areas (and $2\sigma$ uncertainties) for the leading and trailing hemispheres of these moons (Table 11, Figure 9a and 9b). We found a significant difference ($> 2\sigma$) for the 1.52-µm and 2.02-µm $H_2O$ ice bands between the leading and trailing hemispheres of Ariel and Titania. The mean 2.02-µm $H_2O$ ice bands for the leading and trailing hemispheres of Umbriel and Oberon also display significant differences, whereas the 1.52-µm $H_2O$ bands for these two moons does not display significant differences ($> 1\sigma$). We also calculated mean leading/trailing $H_2O$ band area ratios for each moon (Table 12, Figure 9c). These leading/trailing $H_2O$ band area ratios demonstrate that both $H_2O$ ice band complexes are significantly stronger on the leading hemispheres of Ariel, Umbriel, Titania, and Oberon, but not on the innermost classical moon, Miranda.

Thus, both the $F$-test and mean $H_2O$ band area analyses indicate that $H_2O$ ice bands are stronger on the leading hemispheres of Ariel, Umbriel, Titania, and Oberon (statistically significant differences in band areas in most cases). Furthermore, our analyses show that the difference in $H_2O$ band areas between the leading and trailing hemispheres of these moons tends to decrease with distance from Uranus, consistent with enhanced planetocentric dust accumulation on the leading hemispheres of the outer moons.

### *6.4. Latitudinal trends in $H_2O$ band strengths*

Along with mapping longitudinal variations in $H_2O$ band strengths, we compared $H_2O$ ice bands in spectra collected over southern and northern hemispheres for Ariel, Umbriel, Titania, and Oberon. We separated the SpeX spectra into two groups: the southern hemisphere spectra, which were collected by Rivkin and Grundy (sub-observer latitudes ~4 – 30°S) and northern hemisphere spectra, which were gathered by Cartwright (sub-observer latitudes ~17 – 35°N). We further split the southern and northern latitude groups of spectra into leading and trailing quadrants, and then calculated the mean $H_2O$ band areas for all four groups: southern-leading (SL), southern-trailing (ST), northern-leading (NL), and northern-trailing (NT) quadrants (Table 13). Next, we divided the mean leading quadrant $H_2O$ band areas by the mean trailing quadrant $H_2O$ band areas ($H_2O_{SL/ST}$ and $H_2O_{NL/NT}$ for southern and northern latitude spectra, respectively) and compared the resulting ratios (Table 12).

We note subtle differences ($< 2\sigma$) between the $H_2O_{SL/ST}$ and $H_2O_{NL/NT}$ ratios for the mean 1.52-µm and 2.02-µm $H_2O$ ice band areas (Figure 10). The leading/trailing $H_2O$ band area ratios for Ariel, Umbriel, and the 1.52-µm $H_2O$ band area ratio for Titania appear to be larger in the southern latitude dataset compared to the northern latitude dataset. These subtle latitudinal differences in $H_2O$ band areas suggest that more $H_2O$ ice is exposed on the NT quadrants of these moons compared to the ST quadrants. Unlike Ariel, Umbriel, and Titania, the mean $H_2O$ ice band areas for the southern latitude dataset appear to be stronger on the trailing hemisphere of Oberon compared to its leading hemisphere (> 1$\sigma$ difference). This flipped hemispherical asymmetry in the mean $H_2O$ band strengths of Oberon does not hold for the northern latitude spectra of this moon, which have larger mean $H_2O$ band areas on its leading hemisphere, like the other classical moons (> 2$\sigma$ difference). Thus, these moons appear to display only minor variations in $H_2O$ ice band strengths between their southern and northern latitudes. Because we only analyzed one spectrum collected over the southern hemisphere of Miranda, we excluded this moon from our analysis of latitudinal variation in $H_2O$ band strengths on the classical Uranian satellites. We discuss the implications of these subtle latitudinal differences in $H_2O$ ice band areas in greater detail in sections *7.3* and *7.4*.

### *6.5. Analysis of LXD and IRAC albedos and comparison to other icy moons*

We compared broadband L/L' geometric albedos ($A_0$), calculated using the LXD spectra, to broadband IRAC channel 1 geometric albedos, spanning the same wavelength range (~3.1 – 4.0 µm) (for IRAC image reduction and geometric albedo calculation details, see Cartwright et al., 2015). To do this, we convolved the binned versions of the LXD spectra shown in Figure 3 with the IRAC channel 1 filter response curves. The resulting LXD geometric albedos and 1$\sigma$ uncertainties are summarized, and compared to the mean IRAC geometric albedos previously reported in Cartwright et al. (2015), in Table 14. The LXD geometric albedos and their 1$\sigma$ uncertainties are consistent with the mean IRAC geometric albedos and uncertainties for these moons, adding to our confidence in both datasets.

The surfaces of the Uranian moons are dark over VNIR wavelengths ($A_0$ ranging from ~0.23 – 0.56 at ~0.96 µm, Table 1) compared to the surfaces of more $H_2O$ ice-rich moons like Tethys and Rhea (VNIR $A_0$ of 1.23 and 0.949 at 0.55 µm, respectively, Verbiscer et al., 2007). However, the Uranian moons are brighter than the dark material-rich regions of moons like Iapetus and Callisto (VNIR $A_0$ of 0.05 and 0.19 at 0.55 µm, respectively, Dalton et al., 2010, and references therein). The LXD and IRAC albedos indicate that the Uranian moons are brighter than these $H_2O$ ice-rich and dark material-rich moons over L/L' band wavelengths. We provide a visual comparison of the LXD spectrum of Oberon's leading hemisphere to LXD spectra of Rhea and Tethys (Emery et al., 2005), an LXD spectrum of Callisto (gathered by our team in 2015), and a Visual Imaging Mapping Spectrometer (VIMS) spectrum of average dark material on Iapetus (Pinilla-Alonso et al., 2011) (Figure 11).

This comparison demonstrates that ice-rich (Rhea and Tethys) and dark material-rich (Callisto and Iapetus) moons in the Jovian and Saturnian systems have lower albedos in L/L' bands than Oberon. In particular, the peak of the $H_2O$ ice continuum (~3.6 µm) is much more pronounced in the Oberon spectrum compared to these ice-rich and dark material-rich Jovian and Saturnian moons (the other Uranian moons show a similar bright peak in reflectance between ~3.4 and 3.7 µm). Although the albedo of Oberon's spectrum is higher than the ice-rich moons Rhea and Tethys, the shape of the spectral continuum is similar. This 'peaked' continuum shape for Oberon and the other Uranian satellites is distinctly different from the spectra of dark material-rich moons Iapetus and Callisto, which steadily increase in reflectance with increasing wavelength. The similarity of Oberon's continuum shape to Rhea and Tethys in L/L' bands suggests that the Uranian satellites are dominated by the scattering properties of $H_2O$ ice grains over this wavelength range, as opposed to the intimately mixed dark material. The Uranian moons therefore represent a fairly distinct group of icy objects, with relatively dark surfaces in the VIS and NIR (~0.4 – 2.5 µm), and relatively bright surfaces in L/L' bands (~2.9 – 4.1 µm), compared to other $H_2O$ ice-rich moons.

### *6.6. Spectral modeling*

In this section, we present spectral models with which we investigate the composition and grain size range of the red material and other constituents on the Uranian moons. We focus this analysis on spectra of the leading and trailing hemisphere of the outermost moon, Oberon, which is predicted to accumulate the most irregular satellite dust.

#### *6.6.1. Modeling Procedure*

The numerical models were generated using a Hapke-Mie hybrid modeling program. This hybrid approach utilizes Mie theory (*e.g.*, Bohren and Huffman, 1983) to calculate single scattering albedos for each modeled constituent, and then passes these albedos to Hapke equations (*e.g.*, Hapke, 2012) to model the scattering properties of the selected constituents. The Hapke-Mie codes allow us to generate robust numerical models over a wider range of grain sizes than are permissible using Hapke-based scattering codes alone (Moersch and Christensen, 1995). Spectral models that use Mie theory to calculate single scattering albedos can introduce low amplitude resonance artifacts (*i.e.*, small spikes) in certain wavelength ranges, depending on the selected grain sizes of each modeled constituent. To account for these artifacts, our modeling program averages the selected grain sizes over a narrow spread of diameters (typically ~10% spread in grain size for each constituent).

Previous spectral modeling work of Uranian satellite spectra, using the same Hapke-Mie hybrid codes, demonstrated that best fit synthetic spectra between ~1.0 and 2.5 µm provide poor fits to longer wavelength broadband geometric albedos (spanning ~3.1 to 6.5 µm) calculated using Spitzer/IRAC images (Cartwright et al., 2015). Similarly, best fit models that match the IRAC geometric albedos are poor matches to the wavelength region covered by IRTF/SpeX (for more details, see Cartwright et al., 2015).

We generated synthetic spectra for three different wavelength regions of Oberon: 0.7 – 1.3 µm, 1.2 – 2.5 µm, and 2.9 – 4.1 µm, which from hereon we refer to as 'short,' 'mid,' and 'long' wavelength regions, respectively. Because of the relatively low S/N for SXD spectra at shorter wavelengths (< 1.2 µm), and varying wavelength coverage of these data (Table 2), we used the grand average PRISM spectra (shown in Figure 3) to generate best fit models for the short wavelength region. For the mid region, we modeled grand average SXD spectra, and for the long region, we modeled the two Oberon LXD spectra presented here (shown in Figures 3g and 3h).

#### *6.6.2. Modeling Results*

We used intimate mixtures of crystalline $H_2O$ ice (0.2, 10, and 50 µm grain sizes) and amorphous carbon (7 µm grain size) to match the $H_2O$ ice bands in the mid wavelength region, and added areally mixed $CO_2$ ice (5 and 10 μm grain sizes) to the models of Oberon's trailing hemisphere, consistent with previous modeling efforts of this moon (Cartwright et al., 2015). To fit the red-sloped continua of the short wavelength region, we included Triton tholins or amorphous "pyroxene" ($Fs_{40}$, 9 μm grain sizes) in intimate mixtures of crystalline $H_2O$ ice and amorphous C (best fit models summarized in Table 15 and shown in Figure 12).

Our modeling efforts reveal clear differences between the best fit synthetic spectra for the short, mid, and long wavelength ranges. In the short wavelength region, the reflectance levels of our best fit models are sensitive to the abundance and grain size of amorphous carbon, as well as Triton tholins and pyroxene. However, the short wavelength region models are relatively insensitive to changes in $H_2O$ ice abundance, and they do not display the weak $H_2O$ bands centered near 0.81, 0.90, 1.04, and 1.25 μm, consistent with the absence of these shorter $H_2O$ bands in the spectra we have presented. The insensitivity of our short wavelength region models to changing $H_2O$ ice abundance is a result of the small absorption coefficients, and therefore long penetration depths, of $H_2O$ ice over these wavelengths (Mastrapa et al., 2008). The modeled reddening agents, Triton tholins and pyroxene, represent a smaller fraction of our best fit mid wavelength region synthetic spectra, which are dominated by intimate mixtures of $H_2O$ ice and amorphous C. The long wavelength region synthetic spectra are clearly different than the best fit models for either the short or mid wavelength regions, with much smaller grain sizes (primarily 1 μm diameters) and different abundances for all included components. The small grain size-dominated best fit models we generated for the long wavelength region LXD spectra are consistent with previous spectral models, generated using IRAC channel 1 geometric albedos (Cartwright et al., 2015). Although our best fit models fit the shape and albedo of the long wavelength region, they provide poor fits to the possible Fresnel peaks hinted at in the LXD spectra of Titania and Oberon.

Thus, the spectral signature of our selected reddening agents (Triton tholins and amorphous pyroxene) and amorphous C dominate our models in the short wavelength region, $H_2O$ ice and amorphous C dominate our models in the mid region, and $H_2O$ ice dominates the synthetic spectra in the long region. Our short wavelength region models require a larger fraction of reddening agents (pyroxene and Triton tholins) to match the leading hemisphere spectrum of Oberon compared to its trailing hemisphere spectrum. However, the required fraction of reddening agents in the mid and long region models is not as clear, with comparable abundance of pyroxene and Triton tholins required for both the leading and trailing hemisphere of Oberon.

## 7. Discussion

Our reflectance ratio and $H_2O$ ice band area results and analyses indicate that longitudinal and planetocentric trends in composition are present on the classical Uranian satellites, with more spectrally red material on the leading hemispheres of the outer moons. These results and analyses are supported by our short wavelength region best fit models, which require a larger fraction of reddening agents to match the leading hemisphere spectra of Oberon compared to spectra of its trailing hemisphere. In the following sections, we discuss the implications of our results and analyses and compare the Uranian moons to other red, icy objects. We also explore the implications of our spectral modeling work, the different distribution of constituents on Miranda compared to the other classical satellites, and the implications of the possible presence of $NH_3$-hydrates on these moons.

### *7.1. Implications of the observed trends in spectral reddening*

The surfaces of the classical Uranian satellites have likely experienced extensive space weathering from bombardment by charged particles embedded in Uranus' magnetic field (*e.g.*, Stone et al., 1986; Lanzerotti et al., 1987) and collisions with heliocentric dust grains enhanced by gravitational focusing by Uranus (*e.g.*, Zahnle, 2003; Tamayo et al., 2013a). These space weathering processes likely grayed the surfaces of the Uranian moons relatively fast compared to more primitive red objects elsewhere in the outer Solar System (*e.g.*, Luu et al., 1994; Jewitt et al., 1998; Cruikhank et al., 2005), in particular for the inner moons, where charged particle and heliocentric dust impacts are more frequent than on the outer moons. Consequently, the presence of red material on the four largest classical moons indicates recent replenishment of this constituent, most likely due to the accumulation of irregular satellite dust particles generated by recent impacts, analogous to the dust torus around Phoebe that is slowly mantling the leading hemisphere of Iapetus.

Unlike the Voyager 2/ISS results, our analyses indicate that the leading hemisphere of Titania is redder than Oberon, largely due to the reflectance ratios of two spectra with mid-observation longitudes of 21.3º and 29.0º. The mid-observation longitudes of these two spectra are clearly separated from Titania's apex (the location predicted to experience the most heliocentric dust impacts, *e.g.*, Zahnle et al., 2003). Furthermore, the reflectance ratios for these two spectra are significantly larger than the reflectance ratio for the one spectrum collected near Titania's apex (113.8° longitude) (Table 6). Analysis of the Voyager 2/ISS dataset did not lead to the detection of enhanced reddening on Titania (relative to Oberon). It therefore appears that a locally enhanced red spot exists on the northern hemisphere of Titania, between 20-30º longitude. Additionally, slightly weaker $H_2O$ band strengths in spectra collected over the SL quadrant of Oberon and the NL quadrant of Titania coincide with the regions of enhanced reddening on these two moons, demonstrating a possible connection between enhanced reddening and weaker $H_2O$ bands.

Possibly analogous to the red spot we detect on Titania, evidence of reddish smooth regions mantling crater floors are present in the Voyager 2/ISS images of Oberon's southern hemisphere (Smith et al., 1986). These crater floor deposits have been interpreted as cryovolcanically-emplaced units, possibly of organic-rich materials (*e.g.*, Croft and Soderblom, 1991, and references therein). Alternatively, the red material on Titania might have originated in its subsurface and has been recently exposed by tectonic activity, mass wasting, and/or impact events. Consequently, along with mantling by planetocentric red dust, geologic processes could be enhancing the red spectral color of this region of Titania. The absence of spatially-resolved images of the northern hemispheres of the classical Uranian moons limits our ability to search for crater-filling deposits, and other signs of localized geologic activity.

### *7.2. Possible influence of charged particle bombardment on $H_2O$ ice band strengths*

Along with heliocentric and planetocentric dust impacts, bombardment by charged particles trapped in Uranus' magnetic field could contribute to the observed longitudinal asymmetries in $H_2O$ ice band strengths (Cartwright et al., 2015). It has been hypothesized that charged particle radiolysis drives the generation of the detected $CO_2$ ice, converting native $H_2O$ ice and C-rich materials into $CO_2$ (Grundy et al., 2006; Cartwright et al., 2015). This process could enhance leading/trailing asymmetries in $H_2O$ band strengths by reducing the amount of exposed $H_2O$ ice on the trailing hemispheres of these satellites, in particular on the inner moons, where charged particles densities are higher (*e.g.*, Ness et al., 1986, Stone et al., 1986).

The observed differences in SL and NL mean $H_2O$ bands could be the result of enhanced magnetospheric bombardment of either northern or southern latitudes on these moons, driving enhanced generation of $CO_2$ ice and replacing native $H_2O$ ice deposits. The large angular offset (~60°) observed during the Voyager 2 flyby between Uranus' rotational and magnetic field axes (*e.g.*, Ness et al., 1986) appears to qualitatively support preferential charged particle bombardment of the northern hemispheres of these moons. Furthermore, dynamical modeling of charged particle bombardment trajectories demonstrates that low and high energy electrons preferentially impact different latitude and longitude zones of the tidally-locked Galilean moon, Europa (*e.g.*, Cassidy et al., 2013). By analogy to the Jovian system, perhaps low and high energy charged particles in Uranus' magnetosphere also bombard different longitudinal and latitudinal zones of the classical satellites, generating differences in $H_2O$ band strengths on their southern and northern hemispheres. Detailed numerical modeling of magnetic field interactions with the classical Uranian satellites (for example, building on the work of Cao and Paty, 2017) would provide key insight into possible latitudinal preferences of charged particle irradiation and represents critical future work for understanding radiolytic chemistry in the Uranian system.

### *7.3. Implications of the best fit spectral models of Oberon*

The e-folding absorption lengths of photons into $H_2O$ ice are a steep function of wavelength (see Figure 13 in Cartwright et al., 2015), with photons in the short and mid wavelength regions penetrating much further (~0.1 to > 10 mm depths) compared to photons in the long wavelength range (< 0.001 to ~0.05 mm depths) (Mastrapa et al., 2009). These photon penetration depth estimates assume a solid slab with no scattering. Recognizing that scattering off grain boundaries is inherent in a particulate-dominated regolith, Clark and Roush (1984) show that the mean optical path length (MOPL) represents the average distance light travels through a medium without being absorbed, accounting for scattering off internal boundaries. From Hapke (2012, p. 405-406): $MOPL = -1/\alpha * \ln(r)$, where $\alpha$ represents the absorption coefficient and $r$ is reflectance. Utilizing absorption coefficients for crystalline $H_2O$ ice as an input for $\alpha$ and our best fit synthetic spectra of Oberon as the input for $r$, we calculated MOPL ranges of between ~0.15 to > 10 mm for the short and mid wavelength regions and between ~0.001 to 0.05 mm for the long wavelength region. Consequently, comparison of our short/mid and long wavelength region spectral models suggests that we are probing two distinct compositional layers.

Over the short and mid wavelength regions, where $H_2O$ ice absorption is quite weak, photons probe much greater depths into the near-surface of Oberon, and are primarily absorbed by the intimately-mixed dark material. The enhancement of dark material absorption leads to a reduction in albedo over this wavelength range, weakens the 1.52-µm and 2.02-µm $H_2O$ ice bands, and effectively masks the weaker $H_2O$ ice bands between ~0.8 and 1.3 µm. Whereas, over the long wavelength region, where $H_2O$ ice absorbs strongly, the spectra likely probe only the top veneer of Oberon's regolith, where small $H_2O$ ice grains appear to dominate. These small $H_2O$ ice grains enhance backscatter and increase the reflectance of the L/L' band spectra, in particular for the spectral continuum region between ~3.4 and 3.7 µm (on the long wavelength end of the 3-µm $H_2O$ ice band).

### *7.4. Comparing the distribution of constituents on Miranda to the other moons*

Miranda, the innermost classical moon, does not display the same compositional trends as the other classical satellites. The subtly stronger $H_2O$ ice bands on the trailing hemisphere of Miranda might be the result of ring particle bombardment driving enhanced regolith overturn of its trailing hemisphere. In this scenario, ring particles become electrically charged, experience a Lorentz force as magnetic field lines sweep past them, and then slowly spiral outward until their orbits intersect Miranda (Grundy et al., 2006). However, the impact velocities between electrically-charged ring material and Miranda are likely low (Grundy et al., 2006), which would tend to spur mantling of Miranda's trailing hemisphere with dark, non-ice ring material, as opposed to regolith overturn and refreshment of $H_2O$ ice.

$CO_2$ ice has not been detected on the innermost moon, Miranda (Bauer et al., 2002; Grundy et al., 2006; Gourgeot et al., 2014; Cartwright et al., 2015), but the low S/N (~5 – 20) of the available spectra could be obscuring low levels of $CO_2$ ice on this moon ($CO_2$ band depths $\lesssim$ 5% of the continuum). However, even if a small amount of $CO_2$ is present on Miranda, it is clearly less abundant than on the neighboring moon, Ariel, where $CO_2$ band depths are ~10 to 15% of the continuum. It is plausible that the relatively low mass of Miranda (Table 1) allows more $CO_2$ molecules to escape into space, decreasing $CO_2$ molecule retention timescales compared to the other, more massive classical satellites (Grundy et al., 2006).

### *7.5. Possible presence of $NH_3$-hydrates on the Uranian moons*

Some of the spectra presented here display a subtle absorption feature centered near 2.2 µm (Figure 5), similar to the $NH_3$-hydrate feature identified on Charon in ground-based (*e.g.*, Brown and Calvin, 2000) and New Horizons (*e.g.*, Grundy et al., 2016) reflectance spectra. This possible $NH_3$-hydrate absorption feature does not appear to display longitudinal trends (*i.e.*, preferentially on the leading or trailing hemispheres of these moons), suggesting that the distribution of this constituent could be controlled by geologic processes. The potential presence of this constituent on the Uranian moons is a tantalizing possibility given that this volatile can be efficiently destroyed by irradiation (*e.g.*, Strazzulla and Palumbo, 1998; Moore et al., 2007). Thus, $NH_3$-hydrates could be removed by magnetospherically-embedded charged particle irradiation, suggesting that this constituent, if present, was likely emplaced or exposed recently.

Numerous geologic features on these moons, including long linear grooves in chasmata on Ariel and smooth plains and crater deposits on the other moons, have been interpreted as cryovolcanic in origin (*e.g.*, Smith et al., 1986; Jankowski and Squyres, 1988; Schenk, 1991; Kargel, 1994). The subtle 2.2 µm absorption feature could therefore result from cryovolcanic emplacement of $NH_3$-hydrates, and other volatiles, on the surfaces of these satellites. The 2.2 µm absorption feature could also result from exposure of subsurface $NH_3$-hydrate deposits by tectonic processes, which have likely modified the surfaces of these moons (*e.g.*, Smith et al., 1986; Helfenstein et al., 1989; Pappalardo et al., 1997; Beddingfield et al., 2015). Alternatively, $NH_3$-hydrates could be exposed and/or delivered by impactors, possibly analogous to the spatial association between spectra with strong $NH_3$-hydrate features and some of the craters on Charon (Grundy et al., 2016).

## 8. Summary and Future Work

We measured the reflectance ratios ($R_{1.26\,\mu m}/ R_{0.72\,\mu m}$) and 1.52-µm and 2.02-µm $H_2O$ band areas and depths of 84 SpeX spectra to test the hypothesis that the spectrally red material observed on the classical Uranian moons originated on Uranus' retrograde irregular satellites. The reflectance ratios indicate that the leading hemispheres of these moons are redder than their trailing hemispheres, and that the outer moons are redder than the inner moons (subsolar latitudes ~17 to 35°N), consistent with the Voyager 2/ISS spectral colors (subsolar latitude ~81°S). Thus, red material is clearly present at all observed latitudes of the classical Uranian satellites, demonstrating the pervasiveness of this constituent. Furthermore, the fairly homogenous distribution of red material across the leading hemispheres of these moons, and enhanced reddening of the outer moons, is broadly consistent with the accumulation of dust particles that originated on retrograde irregular satellites. Our analysis of $H_2O$ band areas demonstrates that $H_2O$ ice bands are stronger on the leading hemispheres of Ariel, Umbriel, Titania, and Oberon compared to their trailing hemispheres, and leading/trailing longitudinal asymmetries in $H_2O$ band strengths decrease with distance from Uranus. The reduction in longitudinal asymmetries in $H_2O$ ice band strengths with distance from Uranus is consistent with increased accumulation of planetocentric dust on the leading hemispheres of the outer moons. Therefore, our results and analyses support the hypothesis that the spectrally red material, observed in Voyager 2/ISS images and detected in ground-based spectra of the classical moons, originated on retrograde irregular satellites.

We have observed the classical Uranian satellites for the past five years using IRTF/SpeX (subsolar latitudes ~17 – 35ºN). Analysis of the NIR spectra we have collected indicates that the surfaces of these objects are predominantly composed of $H_2O$ ice intimately mixed with a spectrally neutral, dark constituent, with stronger $H_2O$ ice bands on their leading hemispheres. Intimately mixed, spectrally-reddened material is also present, primarily on the leading hemispheres of the outer moons Titania and Oberon. $CO_2$ ice is present primarily on the trailing hemispheres of the inner moons, Ariel and Umbriel, but has not been detected on the innermost classical moon, Miranda. Our relative band area measurements and spectral models indicate that the detected $CO_2$ ice is pure and segregated.

The longitudinal and planetocentric trends in the distribution of red material, $CO_2$ ice, and $H_2O$ ice band strengths suggest that space weathering mechanisms operate differently throughout the Uranian system, thereby modifying the compositions of these moons with different efficiencies and at non-uniform rates. For the inner moons, Ariel and Umbriel, where charged particle irradiation is more intense and heliocentric dust impacts occur more frequently, $CO_2$ ice abundance and longitudinal asymmetries in $H_2O$ ice band strengths are greater. For the outer moons, Titania and Oberon, where planetocentric dust impacts are more frequent, red material is more abundant.

The classical Uranian satellites are located at a heliocentric "crossroads" between the warmer Jovian and Saturnian satellite systems, where the detected $CO_2$ ice is molecularly bound to more refractory components (*e.g.*, Clark et al., 2013, and references therein), and the colder Neptunian satellite, Triton, where $CO_2$ ice is pure and segregated, but displays no longitudinal asymmetries in its distribution and is non-mobile (Holler et al., 2016), unlike the $CO_2$ ice on the Uranian moons. Thus, these moons represent a distinct group of dark, icy objects with different spectral characteristics from the icy moons of the closer in gas giants and icy objects beyond the Uranian system.

New ground- and space-based telescopic observations of the classical Uranian moons, over VIS and NIR wavelengths, would complement and build on the data presented here. Additional observations of Titania could provide key insight into the extent of the localized red spot we have identified on its leading hemisphere. New, high S/N, NIR spectra of Miranda would improve our understanding of the distribution of constituents on the innermost moon. Similarly, additional spectra with high S/N in the 2.1 to 2.3 μm wavelength range would help us search for and characterize the subtle absorption feature centered near 2.2 μm, which hints at the presence of $NH_3$-hydrates on these moons. Investigation of these moons at wavelengths > 2.5 μm is still at an early stage, and additional L/L' band spectra, gathered using ground-based telescopes and NASA's James Webb Space Telescope (scheduled to launch in 2019), would provide us with critical new data to constrain the compositions of the dark material on the Uranian moons. In particular, new, high S/N, L/L' band spectra could determine whether organic species are exposed on the surfaces of these moons by searching for C-H stretching features centered between ~3.3 and 3.4 μm, which could be present in the data we analyzed here but obscured by their low S/N. Additionally, a dedicated mission to the Uranian system would provide critical data for understanding the compositions of these satellites, the origin of their surface constituents, and the system-wide processes that modify their surfaces.

Along with follow up observations, new modeling work of Uranus' plasma environment would provide key insight into interactions between magnetospherically-embedded charged particles and the surfaces of the classical Uranian satellites. These charged particle-satellite surface simulations would dramatically improve our understanding of the resulting radiolytic chemistry on these moons, especially regarding the origin of $CO_2$ ice, and whether the subtle latitudinal asymmetries in $H_2O$ ice band strengths that we have detected are the result of charged particle bombardment. Finally, spectral modeling codes that can accommodate multiple, distinct regolith layers could help future investigations of whether the near-surfaces of the Uranian satellites are compositionally-stratified.

**Acknowledgements**

This study was funded by a NASA Earth and Space Science Fellowship (grant number NNX14AP16H), as well as NASA Solar System Observing grant 16-SSO016_2-0070 and Planetary Astronomy grant NNX10AB23G. Additional funding was provided by Tom Cronin and Helen Sestak whom we thank for their generous support. We also thank William Grundy for access to Uranian satellite SpeX spectra collected by his team between 2001 and 2006. Additionally, we thank the people of Hawaii for allowing us to use Mauna Kea for our observations, as well as the IRTF telescope operators and staff for providing observer support. We thank Daniel Tamayo for his input on dust dynamics in the Uranian system and for providing a review of this manuscript. We also thank an anonymous reviewer for providing insightful comments as well.

**Table 1:** Classical Uranian satellites.

| Satellite | Orbital Radius (km) | Orbital Radius ($R_{Uranus}$) | Orbital Period (days) | Radius (km) | Mass (x $10^{20}$ kg) | Density (g $cm^{-3}$) | *Geo. Albedo ($A_0$) ($\lambda$ ~0.96 μm) |
|---|---|---|---|---|---|---|---|
| Miranda | 129,900 | 5.12 | 1.41 | 236 | 0.66 | 1.21 | 0.45 |
| Ariel | 190,900 | 7.53 | 2.52 | 579 | 13.53 | 1.59 | 0.56 |
| Umbriel | 266,000 | 10.5 | 4.14 | 585 | 11.72 | 1.46 | 0.26 |
| Titania | 436,300 | 17.2 | 8.71 | 789 | 35.27 | 1.66 | 0.39 |
| Oberon | 583,500 | 23.0 | 13.46 | 762 | 30.14 | 1.56 | 0.33 |

**Geometric albedos from Figure 7 in Karkoschka (2001).*

**Table 2:** IRTF/SpeX observing modes.

| Observing PI | Years | SpeX Mode | MORIS? | Wavelength Range (μm) | Slit Width (") | Spectral Resolution ($\lambda/\Delta\lambda$) | Total Number of Spectra | References |
|---|---|---|---|---|---|---|---|---|
| Rivkin | 2000 | SXD | No | ~0.81 - 2.42 | 0.8 | ~750 | 7 | Cartwright et al. (2015) |
| Grundy | 2001 - 2002 | SXD | No | ~0.81 - 2.42 | 0.5 | ~ 1400 | 6 | Grundy et al. (2003) |
| Grundy | 2003 - 2006 | SXD | No | ~0.81 - 2.42 | 0.3 | ~1700 | 15 | Grundy et al. (2006) |
| Gourgeot | 2014 | SXD | No | ~0.81 - 2.42 | 0.8 | ~750 | 2 | Gourgeot et al. (2014) |
| Cartwright | 2012 | SXD | No | ~0.81 - 2.42 | 0.8 | ~750 | 8 | Cartwright et al. (2015) |
| Cartwright | 2013 | SXD | Yes | ~0.94 - 2.42 | 0.8 | ~750 | 5 | Cartwright et al. (2015) |
| Cartwright | 2014 - 2015 | SXD | Yes | ~0.94 - 2.55 | 0.8 | ~750 | 11 | This work |
| Cartwright | 2014 - 2016 | LXD | Yes | ~1.67 - 4.2 | 0.8 | ~938 | 6 | This work |
| Cartwright | 2014 - 2017 | PRISM | No | ~0.70 - 2.52 | 0.8 | ~94 | 30 | This work |

**Table 3:** IRTF/SpeX observations.

| Object | Obser. PI | SpeX Mode | Long. (°) | Lat. (°) | Lead/ Trail | UT Date | UT Time | $t_{int}$ (min) |
|---|---|---|---|---|---|---|---|---|
| Miranda | Rivkin | SXD | 36.3 | -35.4 | Interm. (L) | 9/7/2000 | 9:30 | 37 |
| | Cartwright | PRISM | 92.1 | 30.4 | Leading | 9/12/2015 | 11:10 | 28 |
| | Gourgeot | SXD | 165.0 | 21.4 | Interm. (T) | 9/26/2012 | 10:40 | 64 |
| | Cartwright | PRISM | 236.2 | 30.4 | Trailing | 9/11/2015 | 14:50 | 16 |
| | Gourgeot | SXD | 277.3 | 21.4 | Trailing | 9/25/2012 | 10:30 | 92 |
| | Cartwright | SXD | 279.4 | 24.7 | Trailing | 11/30/2014 | 6:30 | 120 |
| | Cartwright | PRISM | 280.3 | 30.2 | Trailing | 9/17/2015 | 10:40 | 32 |
| Ariel | Cartwright | PRISM | 9.2 | 31.8 | Interm. (L) | 9/18/2015 | 10:05 | 7.5 |
| | Cartwright | SXD | 15.3 | 27.8 | Interm. (L) | 9/15/2014 | 11:35 | 88 |
| | Cartwright | LXD | 28.0 | 31.8 | Interm. (L) | 9/18/2015 | 13:15 | 93.3 |
| | Cartwright | PRISM | 38.8 | 35.8 | Interm. (L) | 9/20/2016 | 14:15 | 7.5 |
| | Grundy | SXD | 53.6 | -16.0 | Leading | 8/9/2003 | 12:15 | 156 |
| | Grundy | SXD | 79.8 | -19.4 | Leading | 7/17/2002 | 13:25 | 108 |
| | Cartwright | SXD | 87.8 | 24.0 | Leading | 9/5/2013 | 11:10 | 92 |
| | Grundy | SXD | 93.5 | -18.1 | Leading | 10/4/2003 | 5:45 | 108 |
| | Cartwright | SXD | 110.1 | 32.0 | Leading | 9/11/2015 | 13:30 | 44 |
| | Cartwright | SXD | 132.2 | 28.5 | Interm. (L) | 8/24/2014 | 14:05 | 40 |
| | Cartwright | PRISM | 137.6 | 34.6 | Interm. (L) | 10/21/2016 | 12:40 | 7.5 |
| | Grundy | SXD | 159.9 | -11.1 | Interm. (L) | 7/15/2004 | 12:00 | 112 |
| | Grundy | SXD | 200.0 | -15.9 | Interm. (T) | 8/5/2003 | 12:00 | 84 |
| | Grundy | SXD | 219.8 | -17.2 | Interm. (T) | 9/7/2003 | 9:35 | 90 |
| | Cartwright | PRISM | 224.8 | 31.8 | Interm. (T) | 9/17/2015 | 9:40 | 10 |
| | Cartwright | PRISM | 233.6 | 32.0 | Trailing | 9/12/2015 | 10:15 | 8 |
| | Cartwright | LXD | 253.0 | 32.0 | Trailing | 9/12/2015 | 13:30 | 66.6 |
| | Cartwright | PRISM | 253.9 | 29.2 | Trailing | 12/2/2015 | 5:25 | 9 |
| | Rivkin | SXD | 257.6 | -29.5 | Trailing | 9/6/2000 | 7:35 | 76 |
| | Cartwright | SXD | 278.3 | 24.8 | Trailing | 8/7/2013 | 13:20 | 44 |
| | Grundy | SXD | 294.8 | -19.3 | Trailing | 7/16/2002 | 13:10 | 140 |
| | Grundy | SXD | 304.8 | -23.2 | Trailing | 7/8/2001 | 14:40 | 48 |
| | Grundy | SXD | 316.6 | -18.2 | Interm. (T) | 10/8/2003 | 7:55 | 132 |
| Umbriel | Cartwright | PRISM | 14.4 | 28.8 | Interm. (L) | 12/2/2015 | 5:50 | 14 |
| | Cartwright | SXD | 38.4 | 20.3 | Interm. (L) | 8/13/2012 | 13:55 | 60 |
| | Rivkin | SXD | 75.2 | -29.3 | Leading | 9/6/2000 | 10:00 | 80 |
| | Cartwright | SXD | 81.7 | 27.4 | Leading | 9/15/2014 | 14:20 | 100 |
| | Grundy | SXD | 92.1 | -11.1 | Leading | 7/16/2004 | 14:25 | 74 |
| | Cartwright | PRISM | 101.6 | 34.2 | Leading | 10/21/2016 | 11:30 | 7.5 |
| | Cartwright | PRISM | 105.9 | 31.7 | Leading | 9/11/2015 | 9:50 | 16 |
| | Cartwright | SXD | 131.3 | 17.5 | Leading | 11/1/2012 | 9:30 | 78 |
| | Cartwright | PRISM | 195.5 | 31.6 | Interm. (T) | 9/12/2015 | 10:35 | 10 |
| | Grundy | SXD | 216.2 | -10.9 | Interm. (T) | 7/5/2004 | 14:10 | 80 |
| | Grundy | SXD | 219.8 | -23.0 | Interm. (T) | 7/7/2001 | 14:10 | 52 |
| | Grundy | SXD | 261.3 | -9.4 | Trailing | 9/18/2005 | 9:21 | 196 |
| | Cartwright | SXD | 264.0 | 23.7 | Trailing | 9/5/2013 | 13:55 | 52 |
| | Cartwright | PRISM | 268.0 | 31.5 | Trailing | 9/17/2015 | 10:00 | 12.5 |
| | Grundy | SXD | 317.6 | -11.4 | Interm. (T) | 7/27/2004 | 11:30 | 184 |
| | Cartwright | PRISM | 323.6 | 32.8 | Interm. (T) | 1/23/2017 | 5:10 | 12 |
| | Cartwright | SXD | 345.8 | 27.5 | Interm. (T) | 9.14.2014 | 11:45 | 100 |
| | Cartwright | PRISM | 355.8 | 31.4 | Interm. (T) | 9/18/2015 | 10:20 | 10 |
| Titania | Cartwright | SXD | 13.6 | 19.0 | Interm. (L) | 9/21/2012 | 11:25 | 32 |
| | Cartwright | PRISM | 21.3 | 28.8 | Interm. (L) | 12/2/2015 | 4:55 | 7.5 |
| | Cartwright | PRISM | 29.0 | 32.7 | Interm. (L) | 1/23/2017 | 6:10 | 7.5 |
| | Cartwright | SXD | 37.2 | 27.4 | Interm. (L) | 9/14/2014 | 14:15 | 60 |
| | Cartwright | SXD | 86.5 | 23.6 | Leading | 9/6/2013 | 10:50 | 60 |
| | Grundy | SXD | 98.0 | -18.1 | Leading | 10/8/2003 | 5:45 | 64 |
| | Cartwright | LXD | 106.5 | 32.2 | Leading | 10/21/2016 | 8:30 | 73.3 |
| | Cartwright | PRISM | 113.8 | 34.2 | Leading | 10/21/2016 | 12:45 | 7.5 |
| | Cartwright | PRISM | 126.2 | 31.4 | Leading | 9/17/2015 | 9:00 | 7.5 |
| | Cartwright | SXD | 160.0 | 18.2 | Interm. (L) | 10/12/2012 | 10:15 | 60 |
| | Cartwright | PRISM | 168.4 | 31.4 | Interm. (L) | 9/18/2015 | 9:30 | 7.5 |
| | Grundy | SXD | 213.9 | -11.1 | Interm. (T) | 7/15/2004 | 14:25 | 64 |
| | Cartwright | SXD | 242.0 | 31.6 | Trailing | 9/11/2015 | 11:30 | 60 |
| | Rivkin | SXD | 258.9 | -29.3 | Trailing | 9/5/2000 | 9:55 | 60 |
| | Cartwright | LXD | 270.3 | 35.4 | Trailing | 9/20/2016 | 11:45 | 80 |
| | Cartwright | PRISM | 274.3 | 35.4 | Trailing | 9/20/2016 | 14:00 | 7.5 |
| | Grundy | SXD | 277.8 | -23.0 | Trailing | 7/7/2001 | 13:10 | 36 |
| | Cartwright | PRISM | 280.8 | 31.6 | Trailing | 9/12/2015 | 9:45 | 8 |
| | Cartwright | SXD | 282.7 | 27.2 | Trailing | 9/20/2014 | 13:00 | 28 |
| | Grundy | SXD | 299.6 | -10.2 | Trailing | 10/13/2005 | 9:05 | 106 |
| | Rivkin | SXD | 342.7 | -29.4 | Interm. (T) | 9/7/2000 | 10:35 | 44 |
| Oberon | Rivkin | SXD | 1.0 | -29.4 | Interm. (L) | 9/7/2000 | 8:30 | 32 |
| | Cartwright | LXD | 26.9 | 27.8 | Interm. (L) | 8/28/2014 | 13:40 | 80 |

| | | | | | | | |
|---|---|---|---|---|---|---|---|
| Cartwright | PRISM | 30.8 | 35.2 | Interm. (L) | 9/20/2016 | 14:30 | 7.5 |
| Cartwright | SXD | 64.8 | 18.1 | Leading | 10/12/2012 | 11:05 | 98 |
| Cartwright | PRISM | 78.0 | 31.5 | Leading | 9/11/2015 | 10:15 | 10 |
| Grundy | SXD | 87.0 | -3.7 | Leading | 8/4/2006 | 13:22 | 92 |
| Cartwright | SXD | 91.0 | 23.7 | Leading | 9/1/2013 | 11:40 | 56 |
| Cartwright | PRISM | 104.5 | 28.6 | Leading | 12/2/2015 | 5:10 | 7.5 |
| Cartwright | PRISM | 104.5 | 31.4 | Leading | 9/12/2015 | 10:00 | 8 |
| Grundy | SXD | 110.6 | -10.2 | Leading | 10/13/2005 | 6:30 | 124 |
| Cartwright | PRISM | 123.2 | 32.6 | Leading | 1/23/2017 | 6:00 | 7.5 |
| Cartwright | PRISM | 131.5 | 24.5 | Leading | 12/18/2014 | 5:05 | 16 |
| Grundy | SXD | 216.2 | -23.1 | Interm. (T) | 7/8/2001 | 13:10 | 48 |
| Cartwright | SXD | 223.7 | 18.9 | Trailing | 9/21/2012 | 10:20 | 64 |
| Cartwright | SXD | 236.1 | 17.4 | Trailing | 11/1/2012 | 6:55 | 40 |
| Cartwright | PRISM | 237.3 | 31.3 | Trailing | 9/17/2015 | 9:25 | 7.5 |
| Cartwright | LXD | 242.3 | 31.3 | Trailing | 9/17/2015 | 13:40 | 66.6 |
| Cartwright | PRISM | 264.6 | 31.2 | Trailing | 9/18/2015 | 9:45 | 7.5 |
| Cartwright | SXD | 266.5 | 20.2 | Trailing | 8/13/2012 | 15:15 | 28 |
| Cartwright | SXD | 279.2 | 27.9 | Trailing | 8/24/2014 | 13:05 | 100 |
| Rivkin | SXD | 307.2 | -29.3 | Trailing | 9/5/2000 | 8:10 | 60 |

**Table 4:** Local standard and solar analog stars.

| Observing PI | UT Date | Solar Analogs |
|---|---|---|
| Rivkin | 9/5/2000 | SAO 164, 16 Cygni A |
| | 9/6/2000 | SAO 164, 16 Cygni Q |
| | 9/7/2000 | SAO 164 |
| Cartwright | 8/13/2012 | SAO 109348, SA 115-271 |
| | 9/21/2012 | SAO 109182, SAO 110201, SA 115-271 |
| | 10/12/2012 | SAO 109182, SA 115-271, Hyades 64 |
| | 11/1/2012 | SAO 109182, SAO 126133, Hyades 64 |
| | 8/7/2013 | SAO 109426, SAO 109450 |
| | 9/1/2013 | SAO 109426, SAO 109450, SAO 109348 |
| | 9/5/2013 | SAO 109426, SAO 109450, SAO 109348 |
| | 9/6/2013 | SAO 109426, SAO 109450, SAO 109348 |
| | 8/24/2014 | SAO 110201, SAO 109567 |
| | 8/28/2014 | SAO 110201, SAO 109567, SAO 109182 |
| | 9/14/2014 | SAO 110201, SAO 109567 |
| | 9/15/2014 | SAO 110201, SAO 109567 |
| | 9/20/2014 | SAO 110201, SAO 109567 |
| | 11/30/2014 | SAO 110201, SAO 109567 |
| | 12/18/2014 | SAO 110201, SAO 109567 |
| | 9/11/2015 | SAO 110201, BD+08 205 |
| | 9/12/2015 | SAO 110201, BD+08 205 |
| | 9/17/2015 | SAO 110201, BD+08 205 |
| | 9/18/2015 | SAO 110201, BD+08 205 |
| | 12/2/2015 | SAO 110201, SAO 109567 |
| | 9/20/2016 | SAO 110201, SAO 146253, SAO 109905 |
| | 10/21/2016 | SAO 110201, SAO 109690 |
| | 1/23/2017 | SAO 110201, SAO 109690 |

**Table 5:** Wavelength range of $H_2O$ ice bands and continua.

| $H_2O$ Band Complex | Short Wave. Continuum (μm) | Long Wave. Continuum (μm) | Band Wave. Range (μm) | Band Width (μm) |
|---|---|---|---|---|
| 1.52-μm | 1.318 - 1.440 | 1.720 - 1.750 | 1.44 - 1.72 | 0.280 |
| 2.02-μm | 1.720 - 1.907 | 2.215 - 2.230 | 1.907 - 2.215 | 0.308 |

**Table 6:** Reflectance ratios ($R_{1.26\ \mu m}$ / $R_{0.72\ \mu m}$).

| Object | Satellite Long. (°) | Reflect. Ratio | Δreflect. Ratio (1$\sigma$) |
|---|---|---|---|
| Ariel | 9.2 | 1.039 | 0.022 |
| | 38.8 | 1.013 | 0.006 |
| | 137.6 | 1.035 | 0.005 |
| | 224.8 | 1.002 | 0.005 |
| | 233.6 | 1.014 | 0.005 |
| | 253.9 | 1.066 | 0.003 |
| Umbriel | 14.4 | 1.064 | 0.006 |
| | 101.6 | 1.062 | 0.008 |
| | 105.9 | 1.084 | 0.007 |
| | 195.5 | 1.077 | 0.006 |
| | 268.0 | 1.053 | 0.005 |
| | 323.6 | 1.021 | 0.006 |
| | 355.8 | 1.075 | 0.006 |
| Titania | 21.3 | 1.142 | 0.003 |
| | 29.0 | 1.134 | 0.006 |
| | 113.8 | 1.075 | 0.008 |
| | 168.4 | 1.093 | 0.005 |
| | 274.3 | 1.060 | 0.003 |
| | 280.8 | 1.085 | 0.011 |
| Oberon | 30.8 | 1.083 | 0.011 |
| | 78.0 | 1.085 | 0.005 |
| | 104.5 | 1.064 | 0.007 |
| | 104.5 | 1.093 | 0.005 |
| | 123.2 | 1.115 | 0.005 |
| | 131.5 | 1.082 | 0.005 |
| | 237.3 | 1.057 | 0.005 |
| | 264.6 | 1.065 | 0.004 |

**Table 7:** $H_2O$ ice band areas and depths.

| | | | Integrated $H_2O$ Band Area ($10^{-2}$ µm) | | | | $H_2O$ Band Depth (µm) | | | |
|---|---|---|---|---|---|---|---|---|---|---|
| Target | Long. (°) | Lat. (°) | 1.52-µm Band | Δarea (1$\sigma$) | 2.02-µm Band | Δarea (1$\sigma$) | 1.52-µm Band | Δdepth (1$\sigma$) | 2.02-µm Band | Δdepth (1$\sigma$) |
| Miranda | 36.3 | -35.4 | 2.62 | 0.02 | 6.61 | 0.15 | 0.169 | 0.005 | 0.360 | 0.007 |
| | 92.1 | 30.4 | 4.43 | 0.06 | 9.06 | 0.15 | 0.256 | 0.011 | 0.501 | 0.051 |
| | 165.0 | 21.4 | 3.32 | 0.79 | 8.00 | 0.39 | 0.226 | 0.015 | 0.433 | 0.040 |
| | 236.2 | 30.4 | 4.24 | 0.12 | 8.90 | 0.24 | 0.243 | 0.023 | 0.517 | 0.044 |
| | 277.3 | 21.4 | 3.32 | 0.79 | 8.07 | 0.22 | 0.226 | 0.015 | 0.433 | 0.040 |
| | 279.4 | 24.7 | 3.34 | 0.06 | 8.04 | 0.10 | 0.209 | 0.014 | 0.451 | 0.022 |
| | 280.3 | 30.2 | 3.89 | 0.05 | 8.13 | 0.08 | 0.229 | 0.010 | 0.446 | 0.025 |
| Ariel | 9.2 | 31.8 | 4.19 | 0.04 | 8.45 | 0.07 | 0.247 | 0.009 | 0.455 | 0.016 |
| | 15.3 | 27.8 | 4.05 | 0.01 | 8.58 | 0.02 | 0.235 | 0.002 | 0.446 | 0.003 |
| | 38.8 | 35.8 | 4.63 | 0.02 | 9.19 | 0.03 | 0.265 | 0.004 | 0.471 | 0.009 |
| | 53.6 | -16.0 | 4.73 | 0.02 | 9.29 | 0.05 | 0.274 | 0.003 | 0.473 | 0.006 |
| | 79.8 | -19.4 | 4.99 | 0.01 | 10.03 | 0.04 | 0.290 | 0.002 | 0.504 | 0.003 |
| | 87.8 | 24.0 | 4.92 | 0.01 | 9.27 | 0.02 | 0.291 | 0.002 | 0.480 | 0.003 |
| | 93.5 | -18.1 | 4.83 | 0.02 | 9.66 | 0.03 | 0.286 | 0.003 | 0.503 | 0.004 |
| | 110.1 | 32.0 | 4.68 | 0.01 | 8.95 | 0.02 | 0.275 | 0.002 | 0.481 | 0.004 |
| | 132.2 | 28.5 | 4.31 | 0.01 | 8.44 | 0.02 | 0.258 | 0.002 | 0.441 | 0.005 |
| | 137.6 | 34.6 | 3.96 | 0.03 | 8.59 | 0.04 | 0.234 | 0.010 | 0.435 | 0.009 |
| | 159.9 | -11.1 | 3.23 | 0.02 | 8.57 | 0.03 | 0.202 | 0.005 | 0.448 | 0.005 |
| | 200.0 | -15.9 | 2.98 | 0.02 | 7.48 | 0.12 | 0.182 | 0.003 | 0.395 | 0.006 |
| | 219.8 | -17.2 | 2.67 | 0.01 | 6.84 | 0.02 | 0.166 | 0.002 | 0.360 | 0.004 |
| | 224.8 | 31.8 | 3.18 | 0.01 | 7.30 | 0.02 | 0.195 | 0.002 | 0.394 | 0.008 |
| | 233.6 | 32.0 | 3.16 | 0.02 | 7.01 | 0.05 | 0.196 | 0.003 | 0.374 | 0.010 |
| | 253.9 | 29.2 | 2.88 | 0.02 | 7.38 | 0.02 | 0.187 | 0.004 | 0.400 | 0.008 |
| | 257.6 | -29.5 | 3.25 | 0.03 | 7.18 | 0.09 | 0.188 | 0.005 | 0.375 | 0.005 |
| | 278.3 | 24.8 | 2.90 | 0.02 | 7.15 | 0.04 | 0.177 | 0.005 | 0.383 | 0.010 |
| | 294.8 | -19.3 | 2.90 | 0.02 | 6.74 | 0.04 | 0.177 | 0.002 | 0.388 | 0.004 |
| | 304.8 | -23.2 | 3.02 | 0.03 | 6.55 | 0.19 | 0.184 | 0.004 | 0.384 | 0.007 |
| | 316.6 | -18.2 | 3.21 | 0.02 | 7.19 | 0.03 | 0.194 | 0.004 | 0.387 | 0.008 |
| Umbriel | 14.4 | 28.8 | 1.69 | 0.03 | 5.04 | 0.04 | 0.108 | 0.005 | 0.296 | 0.012 |
| | 38.4 | 20.3 | 1.74 | 0.02 | 4.64 | 0.03 | 0.116 | 0.004 | 0.274 | 0.005 |
| | 75.2 | -29.3 | 1.87 | 0.03 | 5.05 | 0.17 | 0.120 | 0.006 | 0.285 | 0.007 |
| | 81.7 | 27.4 | 1.43 | 0.01 | 4.59 | 0.05 | 0.092 | 0.003 | 0.272 | 0.007 |
| | 92.1 | -11.1 | 1.48 | 0.04 | 4.50 | 0.06 | 0.097 | 0.005 | 0.291 | 0.012 |
| | 101.6 | 34.2 | 1.82 | 0.04 | 4.85 | 0.06 | 0.118 | 0.009 | 0.277 | 0.021 |
| | 105.9 | 31.7 | 2.05 | 0.03 | 5.22 | 0.10 | 0.122 | 0.006 | 0.274 | 0.015 |
| | 131.3 | 17.5 | 1.57 | 0.01 | 4.74 | 0.02 | 0.102 | 0.003 | 0.276 | 0.004 |
| | 195.5 | 31.6 | 1.75 | 0.02 | 4.60 | 0.05 | 0.107 | 0.009 | 0.237 | 0.011 |
| | 216.2 | -10.9 | 1.42 | 0.04 | 4.31 | 0.05 | 0.091 | 0.006 | 0.266 | 0.011 |
| | 219.8 | -23.0 | 1.68 | 0.03 | 3.36 | 0.10 | 0.114 | 0.005 | 0.270 | 0.010 |
| | 261.3 | -9.4 | 0.87 | 0.03 | 3.99 | 0.18 | 0.057 | 0.004 | 0.232 | 0.005 |
| | 264.0 | 23.7 | 1.26 | 0.01 | 4.22 | 0.01 | 0.093 | 0.002 | 0.238 | 0.003 |
| | 268.0 | 31.5 | 1.54 | 0.03 | 4.40 | 0.04 | 0.096 | 0.006 | 0.264 | 0.016 |
| | 317.6 | -11.4 | 1.04 | 0.02 | 4.18 | 0.04 | 0.071 | 0.004 | 0.240 | 0.007 |
| | 323.6 | 32.8 | 1.53 | 0.03 | 4.43 | 0.05 | 0.100 | 0.006 | 0.245 | 0.010 |
| | 345.8 | 27.5 | 1.31 | 0.01 | 4.35 | 0.02 | 0.086 | 0.002 | 0.249 | 0.005 |
| | 355.8 | 31.4 | 1.73 | 0.02 | 4.46 | 0.05 | 0.118 | 0.004 | 0.253 | 0.010 |
| Titania | 13.6 | 19.0 | 3.67 | 0.01 | 8.41 | 0.01 | 0.221 | 0.002 | 0.452 | 0.002 |
| | 21.3 | 28.8 | 4.14 | 0.01 | 9.24 | 0.02 | 0.257 | 0.006 | 0.481 | 0.005 |
| | 29.0 | 32.7 | 4.47 | 0.02 | 9.49 | 0.03 | 0.258 | 0.007 | 0.500 | 0.009 |
| | 37.2 | 27.4 | 4.16 | 0.01 | 8.85 | 0.01 | 0.249 | 0.002 | 0.470 | 0.003 |
| | 86.5 | 23.6 | 4.43 | 0.01 | 9.07 | 0.01 | 0.265 | 0.001 | 0.477 | 0.002 |
| | 98.0 | -18.1 | 4.44 | 0.03 | 9.05 | 0.05 | 0.270 | 0.004 | 0.484 | 0.007 |
| | 113.8 | 34.2 | 4.06 | 0.02 | 8.80 | 0.03 | 0.240 | 0.004 | 0.474 | 0.006 |
| | 126.2 | 31.4 | 4.26 | 0.01 | 8.92 | 0.02 | 0.257 | 0.005 | 0.466 | 0.005 |
| | 160.0 | 18.2 | 3.63 | 0.01 | 8.24 | 0.02 | 0.223 | 0.003 | 0.449 | 0.003 |
| | 168.4 | 31.4 | 4.00 | 0.02 | 8.27 | 0.02 | 0.242 | 0.004 | 0.434 | 0.006 |
| | 213.9 | -11.1 | 3.31 | 0.02 | 8.17 | 0.03 | 0.203 | 0.004 | 0.441 | 0.004 |
| | 242.0 | 31.6 | 3.77 | 0.01 | 6.70 | 0.03 | 0.240 | 0.002 | 0.407 | 0.003 |
| | 258.9 | -29.3 | 3.46 | 0.02 | 8.04 | 0.04 | 0.217 | 0.003 | 0.431 | 0.003 |
| | 274.3 | 35.4 | 3.67 | 0.02 | 7.89 | 0.04 | 0.223 | 0.002 | 0.420 | 0.013 |
| | 277.8 | -23.0 | 3.70 | 0.02 | 8.10 | 0.04 | 0.225 | 0.003 | 0.443 | 0.004 |
| | 280.8 | 31.6 | 3.66 | 0.04 | 8.27 | 0.09 | 0.223 | 0.007 | 0.442 | 0.017 |
| | 282.7 | 27.2 | 3.50 | 0.01 | 7.53 | 0.03 | 0.215 | 0.002 | 0.414 | 0.003 |
| | 299.6 | -10.2 | 3.35 | 0.02 | 8.05 | 0.02 | 0.207 | 0.003 | 0.434 | 0.004 |
| | 342.7 | -29.4 | 3.83 | 0.02 | 9.01 | 0.29 | 0.231 | 0.002 | 0.452 | 0.004 |
| Oberon | 1.0 | -29.4 | 2.56 | 0.02 | 7.18 | 0.32 | 0.162 | 0.003 | 0.383 | 0.007 |
| | 30.8 | 35.2 | 2.86 | 0.06 | 6.83 | 0.06 | 0.167 | 0.012 | 0.389 | 0.015 |
| | 64.8 | 18.1 | 2.63 | 0.01 | 6.52 | 0.02 | 0.172 | 0.003 | 0.363 | 0.004 |
| | 78.0 | 31.5 | 3.06 | 0.01 | 6.90 | 0.03 | 0.193 | 0.004 | 0.381 | 0.008 |
| | 87.0 | -3.7 | 2.73 | 0.05 | 6.29 | 0.16 | 0.178 | 0.006 | 0.375 | 0.010 |
| | 91.0 | 23.7 | 2.90 | 0.03 | 6.78 | 0.04 | 0.181 | 0.006 | 0.353 | 0.007 |

| | | | | | | | | | |
|---|---|---|---|---|---|---|---|---|---|
| 104.5 | 28.6 | 2.84 | 0.02 | 6.80 | 0.03 | 0.179 | 0.006 | 0.371 | 0.009 |
| 104.5 | 31.4 | 2.86 | 0.01 | 7.11 | 0.02 | 0.185 | 0.005 | 0.383 | 0.005 |
| 110.6 | -10.2 | 2.16 | 0.02 | 5.71 | 0.02 | 0.140 | 0.003 | 0.318 | 0.004 |
| 123.2 | 32.6 | 3.15 | 0.02 | 7.18 | 0.03 | 0.181 | 0.006 | 0.384 | 0.007 |
| 131.5 | 24.5 | 2.87 | 0.02 | 6.85 | 0.03 | 0.188 | 0.004 | 0.376 | 0.010 |
| 216.2 | -23.1 | 2.75 | 0.02 | 6.12 | 0.06 | 0.173 | 0.003 | 0.375 | 0.005 |
| 223.7 | 18.9 | 2.21 | 0.01 | 6.09 | 0.01 | 0.143 | 0.002 | 0.338 | 0.003 |
| 236.1 | 17.4 | 2.31 | 0.01 | 6.30 | 0.01 | 0.150 | 0.002 | 0.350 | 0.003 |
| 237.3 | 31.3 | 2.81 | 0.02 | 6.35 | 0.03 | 0.169 | 0.004 | 0.342 | 0.007 |
| 264.6 | 31.2 | 2.76 | 0.01 | 6.32 | 0.03 | 0.177 | 0.003 | 0.345 | 0.006 |
| 266.5 | 20.2 | 2.46 | 0.02 | 6.07 | 0.03 | 0.157 | 0.005 | 0.335 | 0.008 |
| 279.2 | 27.9 | 2.67 | 0.01 | 6.10 | 0.01 | 0.159 | 0.002 | 0.336 | 0.003 |
| 307.2 | -29.3 | 2.46 | 0.01 | 6.35 | 0.04 | 0.157 | 0.002 | 0.347 | 0.002 |

**Table 8:** *F*-test analysis of reflectance ratios and $H_2O$ ice band areas.

| Satellite | Analyzed Measurement | One Tailed *F*-test Ratio | Sample Size (n) | Mean Model Deg. Freedom (n - 1) | Sinusoidal Model Deg. Freedom (n - 3) | Probability (*p*) | Reject Null Hypothesis? |
|---|---|---|---|---|---|---|---|
| Miranda | Reflect. ratio | - | - | - | - | - | - |
| | 1.52-µm $H_2O$ | 44.274 | 7 | 6 | 4 | $1.3*10^{-3}$ | Yes |
| | 2.02-µm $H_2O$ | 3.642 | 7 | 6 | 4 | $1.16*10^{-1}$ | No |
| Ariel | Reflect. ratio | 0.436 | 6 | 5 | 3 | $8.05*10^{-1}$ | No |
| | 1.52-µm $H_2O$ | 134.243 | 21 | 20 | 18 | << 0.0001 | Yes |
| | 2.02-µm $H_2O$ | 89.713 | 21 | 20 | 18 | << 0.0001 | Yes |
| Umbriel | Reflect ratio | 1.474 | 7 | 6 | 4 | $3.69*10^{-1}$ | No |
| | 1.52-µm $H_2O$ | 5.471 | 18 | 17 | 15 | $9.31*10^{-4}$ | Yes |
| | 2.02-µm $H_2O$ | 23.845 | 18 | 17 | 15 | << 0.0001 | Yes |
| Titania | Reflect ratio | 11.619 | 6 | 5 | 3 | $3.53*10^{-2}$ | Yes |
| | 1.52-µm $H_2O$ | 14.241 | 19 | 18 | 16 | << 0.0001 | Yes |
| | 2.02-µm $H_2O$ | 25.928 | 19 | 18 | 16 | << 0.0001 | Yes |
| Oberon | Reflect ratio | 33.959 | 8 | 7 | 5 | $6.38*10^{-4}$ | Yes |
| | 1.52-µm $H_2O$ | 6.399 | 19 | 18 | 16 | $2.53*10^{-4}$ | Yes |
| | 2.02-µm $H_2O$ | 8.939 | 19 | 18 | 16 | $2.95*10^{-5}$ | Yes |

**Table 9:** Mean reflectance ratios ($R_{1.26\ \mu m} / R_{0.72\ \mu m}$).

| Object | Number of PRISM Spec. | Leading/ Trailing | Reflect. Ratio | ΔReflect. Ratio (2$\sigma$) | Gradient per 0.1 µm (%) | *ΔGradient per 0.1 µm (%) |
|---|---|---|---|---|---|---|
| Ariel | 3 | L | 1.029 | 0.016 | 0.500 | 0.052 |
| | 3 | T | 1.029 | 0.011 | 0.586 | 0.032 |
| Umbriel | 3 | L | 1.070 | 0.010 | 1.133 | 0.036 |
| | 4 | T | 1.058 | 0.010 | 1.086 | 0.030 |
| Titania | 4 | L | 1.118 | 0.010 | 2.232 | 0.027 |
| | 2 | T | 1.072 | 0.012 | 1.332 | 0.043 |
| Oberon | 6 | L | 1.092 | 0.008 | 1.664 | 0.025 |
| | 2 | T | 1.064 | 0.008 | 1.049 | 0.022 |

**Spectral gradient errors only represent slope fitting uncertainties and do not include observational errors (i.e., accounting for variations in solar analog slope and temporal changes in the observing conditions each night).*

**Table 10:**

| Object | PRISM ($R_{1.26\ \mu m}/R_{0.72\ \mu m}$) | | ISS Gr/Vi Colors ($R_{0.55\ \mu m}/R_{0.41\ \mu m}$) | | |
|---|---|---|---|---|---|
| | Lead./Trail. Reflect. Ratios | Δlead./Trail. Reflect. Ratios (%) | *Average Color Ratios | *Lead./Trail. Color Ratios | *ΔColor Ratios (%) |
| Ariel | 1.00 | 1.8 | 1.07 | 1.02 | 2 |
| Umbriel | 1.01 | 1.3 | 1.05 | 1.08 | 3 |
| Titania | 1.04 | 1.5 | 1.12 | 1.08 | 2 |
| Oberon | 1.03 | 1.0 | 1.15 | 1.23 | 2 |

**Values reported in Table 4 of Buratti and Mosher (1991).*

**Table 11:** Mean leading and trailing $H_2O$ ice band areas.

| | | | Integrated $H_2O$ Band Area ($10^{-2}$ µm) | | | |
|---|---|---|---|---|---|---|
| Satellite | Leading/ Trailing | Number of Spectra | 1.52-µm Band | Δarea ($2\sigma$) | 2.02-µm Band | Δarea ($2\sigma$) |
| Miranda | L | 3 | 3.46 | 1.18 | 7.89 | 1.45 |
| | T | 4 | 3.70 | 0.70 | 8.29 | 0.45 |
| Ariel | L | 11 | 4.41 | 0.32 | 9.00 | 0.32 |
| | T | 10 | 3.02 | 0.12 | 7.08 | 0.19 |
| Umbriel | L | 8 | 1.71 | 0.15 | 4.83 | 0.19 |
| | T | 10 | 1.41 | 0.19 | 4.23 | 0.23 |
| Titania | L | 10 | 4.13 | 0.19 | 8.81 | 0.29 |
| | T | 9 | 3.58 | 0.12 | 7.97 | 0.42 |
| Oberon | L | 11 | 2.78 | 0.16 | 6.74 | 0.27 |
| | T | 8 | 2.55 | 0.16 | 6.21 | 0.09 |

**Table 12:** Mean leading/trailing $H_2O$ band area ratios.

| | | All Latitudes | | Southern Latitudes | | Northern Latitudes | |
|---|---|---|---|---|---|---|---|
| Satellite | $H_2O$ Band | Lead./Trail. Band Area Ratio | Δband Ratio ($2\sigma$) | Lead//Trail. Band Area Ratio | Δband Ratio ($2\sigma$) | Lead./Trail. Band Area Ratio | Δband Ratio ($2\sigma$) |
| Miranda | 1.52-µm | 0.94 | 0.36 | - | - | - | - |
| | 2.02-µm | 0.95 | 0.18 | - | - | - | - |
| Ariel | 1.52-µm | 1.46 | 0.12 | 1.48 | 0.285 | 1.45 | 0.119 |
| | 2.02-µm | 1.27 | 0.06 | 1.34 | 0.105 | 1.22 | 0.047 |
| Umbriel | 1.52-µm | 1.21 | 0.192 | 1.34 | 0.502 | 1.21 | 0.197 |
| | 2.02-µm | 1.14 | 0.076 | 1.13 | 0.17 | 1.11 | 0.052 |
| Titania | 1.52-µm | 1.15 | 0.066 | 1.26 | 0.074 | 1.13 | 0.065 |
| | 2.02-µm | 1.11 | 0.068 | 1.09 | 0.053 | 1.16 | 0.109 |
| Oberon | 1.52-µm | 1.09 | 0.093 | 0.953 | 0.168 | 1.142 | 0.101 |
| | 2.02-µm | 1.09 | 0.046 | 1.025 | 0.148 | 1.089 | 0.035 |

**Table 13:** Mean leading and trailing $H_2O$ band areas at southern and northern latitudes.

| | | | | Integrated $H_2O$ Band Area ($10^{-2}$ μm) | | | |
|---|---|---|---|---|---|---|---|
| Satellite | Leading/ Trailing | Southern/ Northern | Number of Spectra | 1.52-μm Band | Δarea (2σ) | 2.02-μm Band | Δarea (2σ) |
| Ariel | L | S | 4 | 4.45 | 0.82 | 9.39 | 0.62 |
| | | N | 7 | 4.39 | 0.27 | 8.78 | 0.27 |
| | T | S | 6 | 3.01 | 0.18 | 7.00 | 0.29 |
| | | N | 4 | 3.03 | 0.16 | 7.21 | 0.17 |
| Umbriel | L | S | 2 | 1.68 | 0.39 | 4.78 | 0.58 |
| | | N | 6 | 1.72 | 0.18 | 4.85 | 0.20 |
| | T | S | 4 | 1.25 | 0.37 | 3.96 | 0.44 |
| | | N | 6 | 1.52 | 0.17 | 4.37 | 0.09 |
| Titania | L | S | 1 | 4.44 | 0.06 | 9.05 | 0.10 |
| | | N | 9 | 4.09 | 0.20 | 8.81 | 0.29 |
| | T | S | 5 | 3.53 | 0.20 | 8.27 | 0.39 |
| | | N | 4 | 3.61 | 0.11 | 7.60 | 0.67 |
| Oberon | L | S | 3 | 2.48 | 0.34 | 6.39 | 0.89 |
| | | N | 8 | 2.90 | 0.11 | 6.76 | 0.18 |
| | T | S | 2 | 2.61 | 0.29 | 6.24 | 0.24 |
| | | N | 6 | 2.54 | 0.20 | 6.21 | 0.11 |

**Table 14:** Mean leading and trailing Spitzer/IRAC and SpeX/LXD geometric albedos.

| | | Spitzer/IRAC | | SpeX/LXD | |
|---|---|---|---|---|---|
| Target | Leading/ Trailing | Mean Geometric Albedo | ΔMean Geometric Albedo ($1\sigma$) | Geometric Albedo | ΔGeometric Albedo ($1\sigma$) |
| Ariel | L | 0.188 | 0.004 | 0.203 | 0.005 |
| | T | 0.232 | 0.005 | 0.229 | 0.007 |
| Umbriel | L | 0.153 | 0.006 | - | - |
| | T | 0.158 | 0.002 | - | - |
| Titania | L | 0.160 | 0.002 | 0.148 | 0.002 |
| | T | 0.160 | 0.003 | 0.162 | 0.002 |
| Oberon | L | 0.167 | 0.002 | 0.166 | 0.003 |
| | T | 0.168 | 0.001 | 0.167 | 0.003 |

**Table 15:** Best fit synthetic spectra of Oberon.

| | "Short" Model (0.6 - 1.3 µm) | | "Mid" Model (1.3 - 2.5 µm) | | "Long" Model (2.9 - 4.2 µm) | |
|---|---|---|---|---|---|---|
| Leading/Trailing | Model Components | Mix (%) | Model Components | Mix (%) | Model Components | Mix (%) |
| Oberon Leading | 50 µm $H_2O$ | 23.4 | 50 µm $H_2O$ | 26.0 | 10 µm $H_2O$ | 14.0 |
| Organic-rich model | 10 µm $H_2O$ | 31.7 | 10 µm $H_2O$ | 35.7 | 1 µm $H_2O$ | 60.0 |
| | 0.2 µm $H_2O$ | 0.3 | 0.2 µm $H_2O$ | 0.3 | 0.2 µm $H_2O$ | 1.0 |
| | 7 µm amorphous C | 17.6 | 7 µm amorphous C | 21.0 | 1 µm amorphous C | 15.0 |
| | 9 µm Triton tholin | 27.0 | 9 µm Triton tholin | 17.0 | 1 µm Triton tholin | 10.0 |
| Silicate-rich model | 50 µm $H_2O$ | 24.0 | 50 µm $H_2O$ | 19.9 | 10 µm $H_2O$ | 14.0 |
| | 10 µm $H_2O$ | 33.7 | 10 µm $H_2O$ | 41.0 | 1 µm $H_2O$ | 60.0 |
| | 0.2 µm $H_2O$ | 0.3 | 0.2 µm $H_2O$ | 0.3 | 0.2 µm $H_2O$ | 1.0 |
| | 7 µm amorphous C | 15.0 | 7 µm amorphous C | 18.7 | 1 µm amorphous C | 15.0 |
| | 9 µm pyroxene | 27.0 | 9 µm pyroxene | 20.1 | 1 µm pyroxene | 10.0 |
| *Oberon Trailing | 50 µm $H_2O$ | 25.0 | 50 µm $H_2O$ | 30.0 | 10 µm $H_2O$ | 26.0 |
| Organic-rich model | 10 µm $H_2O$ | 33.7 | 10 µm $H_2O$ | 29.7 | 1 µm $H_2O$ | 47.0 |
| | 0.2 µm $H_2O$ | 0.3 | 0.2 µm $H_2O$ | 0.3 | 0.2 µm $H_2O$ | 1.0 |
| | 7 µm amorphous C | 21.0 | 7 µm amorphous C | 20.5 | 1 µm amorphous C | 16.0 |
| | 9 µm Triton tholin | 20.0 | 9 µm Triton tholin | 19.5 | 1 µm Triton tholins | 10.0 |
| *Oberon Trailing | 50 µm $H_2O$ | 27.0 | 50 µm $H_2O$ | 22.0 | 10 µm $H_2O$ | 16.0 |
| Silicate-rich model | 10 µm $H_2O$ | 33.7 | 10 µm $H_2O$ | 38.7 | 1 µm $H_2O$ | 53.0 |
| | 0.2 µm $H_2O$ | 0.3 | 0.2 µm $H_2O$ | 0.3 | 0.2 µm $H_2O$ | 0.8 |
| | 7 µm amorphous C | 17.0 | 7 µm amorphous C | 20.0 | 1 µm amorphous C | 17.0 |
| | 9 µm pyroxene | 22.0 | 9 µm pyroxene | 19.0 | 1 µm pyroxene | 13.2 |

**Oberon trailing models are intimate mixtures, scaled to 97%, with 3% pure $CO_2$ ice added linearly to represent areally segregated deposits (see Appendix B in Cartwright et al., 2015).*

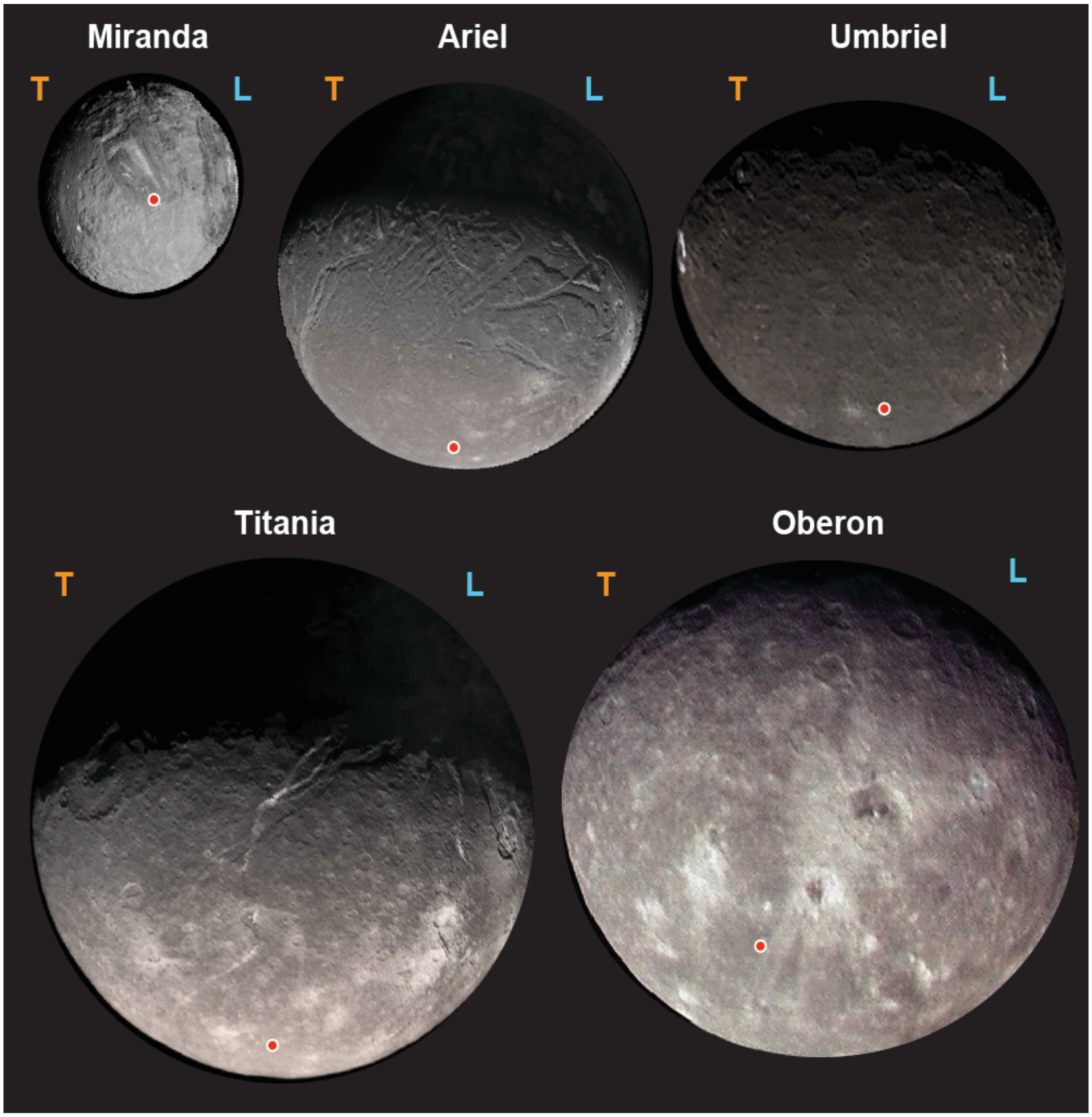

**Figure 1:** Voyager 2/ISS image mosaics of the classical Uranian satellites (courtesy NASA/JPL/-Caltech/USGS), with night-side sections of Ariel and Titania revealed using 'Uranus shine' (Stryk and Stooke, 2008), qualitatively scaled to their relative sizes (radius and mass of each moon listed in Table 1). Imaged sections of each moon's leading (blue L) and trailing (orange T) hemispheres, and the approximate location of their south poles (red circle), are shown as well (subsolar point at the time of Voyager 2 encounter with Uranus, ~81°S). Spatial resolution ranges from a few hundred m/pixel for Miranda to ~6 km/pixel for Oberon. Figure 2 shows map-projected versions of these Voyager 2/ISS image mosaics, which include other lower quality images not shown here, in particular for the leading hemisphere of Ariel.

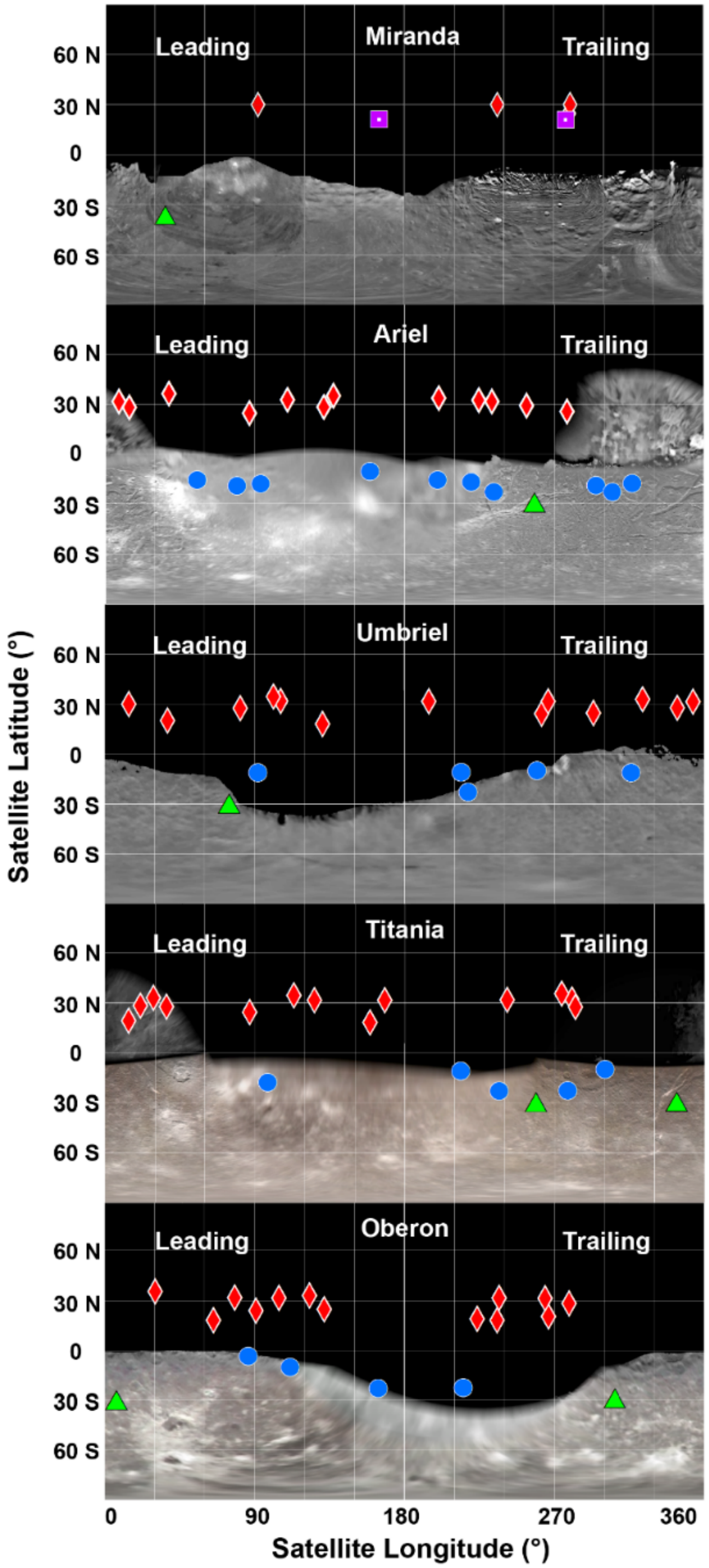

**Figure 2:** Mid-observation satellite longitudes and latitudes of the classical Uranian satellites, made by three different teams using IRTF/SpeX: Red diamonds (PI Cartwright, 2012 – 2017), blue circles (PI Grundy, 2001 – 2006), green triangles (PI Rivkin, 2000), and purple squares (PI Gourgeot, 2012). Base maps are Voyager 2/ISS image mosaics (courtesy NASA/JPL/Caltech/USGS, http://maps.jpl.nasa.gov/uranus.html). Night-side sections of Ariel and Titania are revealed in 'Uranus shine' (Stryk and Stooke, 2008).

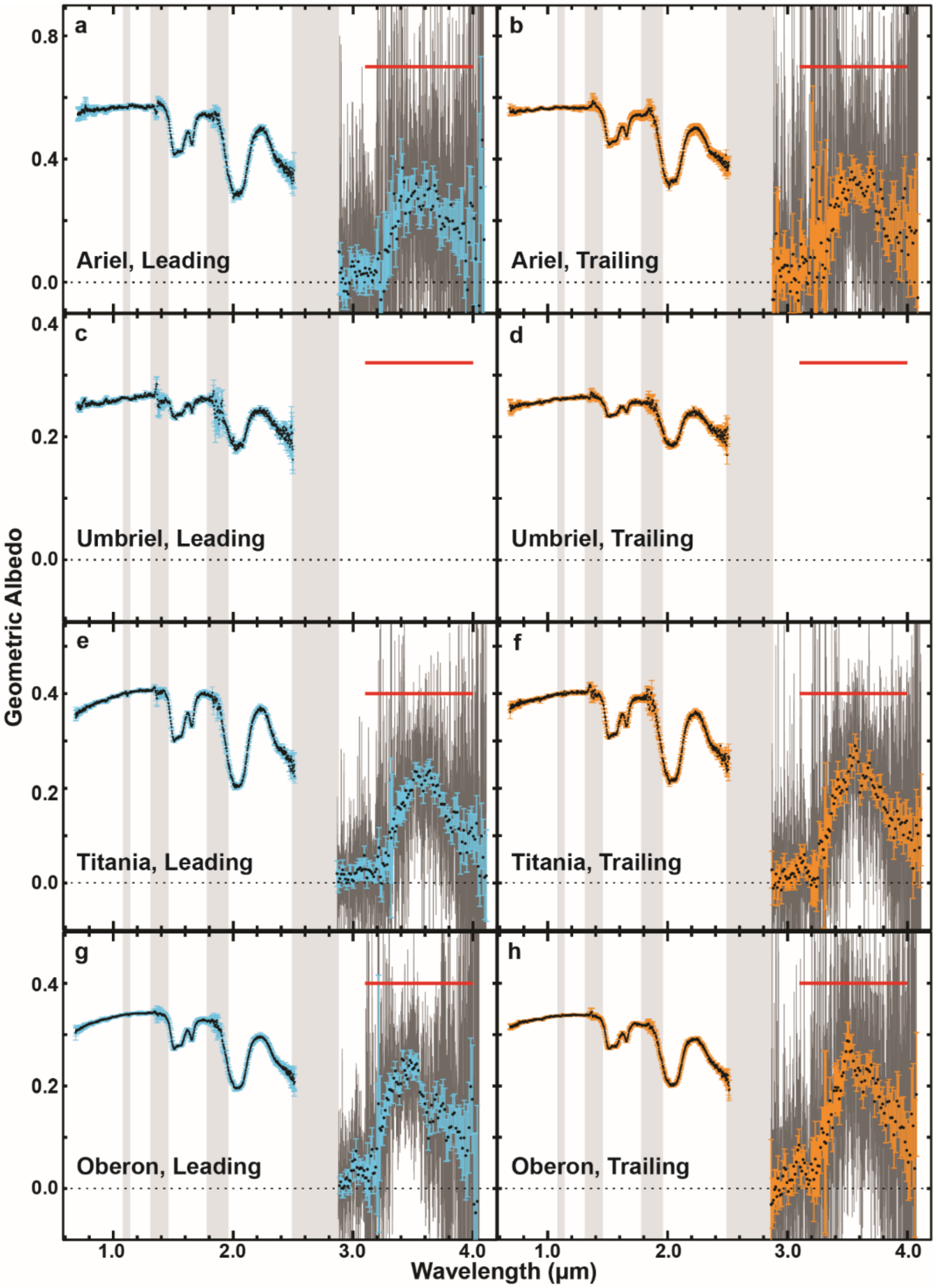

**Figure 3:** 'Grand average' PRISM mode SpeX spectra and the six 'native-resolution' LXD mode SpeX spectra we have gathered (dark gray), along with 'binned' versions of the LXD spectra (50 pixel wide binning window). All spectra shown are scaled to geometric albedo at ~0.96 μm (summarized in Table 1). (a, c, e, g) Leading hemisphere spectra (blue 1$\sigma$ error bars) and (b, d, f, h) trailing hemisphere spectra (orange 1$\sigma$ error bars) are displayed for each moon. Wavelength range of telluric bands (light gray regions) and IRAC channel 1 (~3.1 – 4.0 μm, red bars) are indicated.

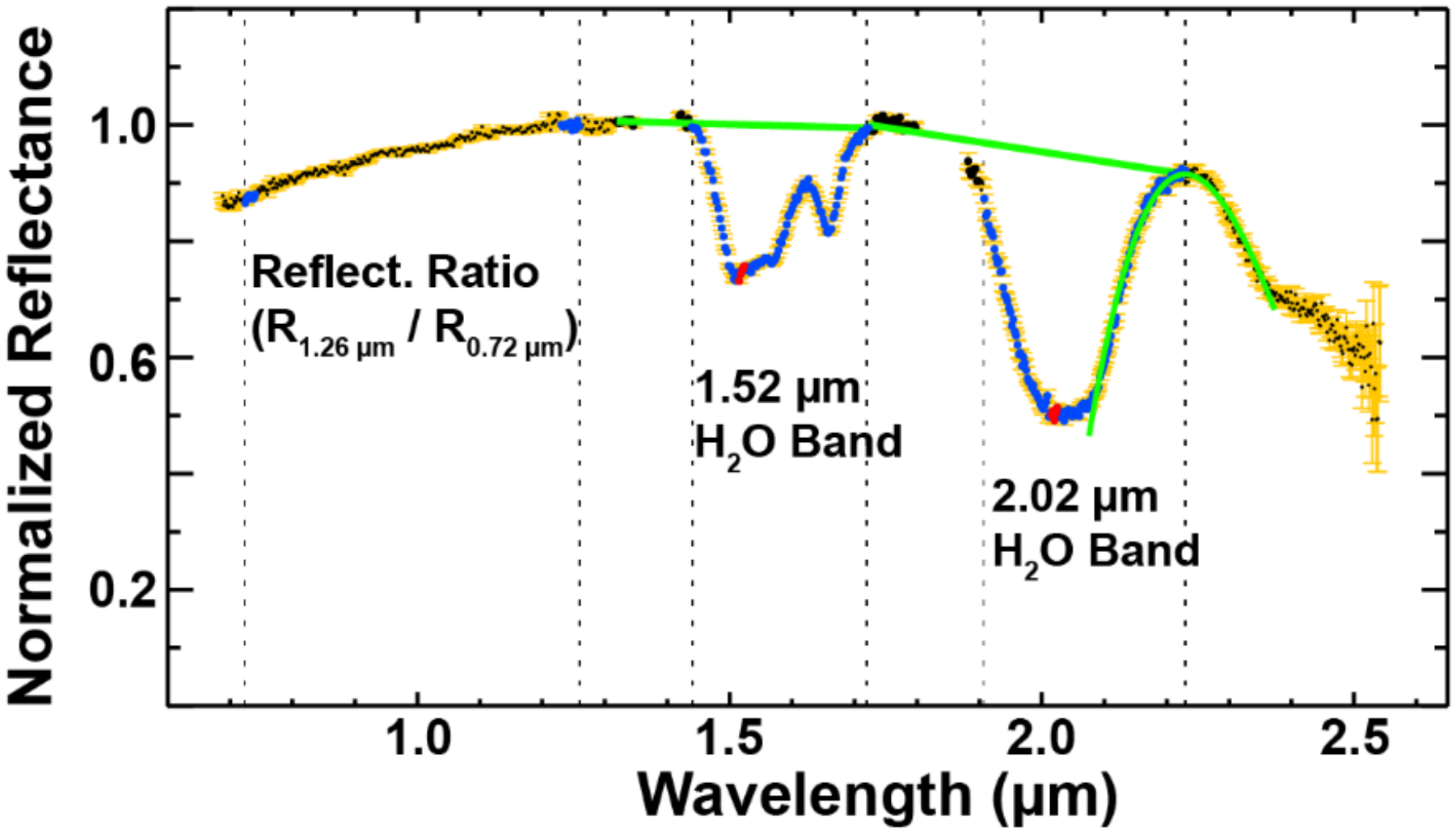

**Figure 4:** Examples of spectral measurements conducted on a SpeX spectrum (PRISM mode) gathered over Titania's leading hemisphere (mid observation satellite longitude 21°). Data points used in reflectance ratio ($R_{1.26\ \mu m}$ / $R_{0.72\ \mu m}$) and $H_2O$ band area measurements (blue) and $H_2O$ band center measurements (red) are highlighted. Continua for $H_2O$ band area measurements are also shown, including third order polynomial utilized to identify long wavelength end of 2.02-$H_2O$ µm band (green lines).

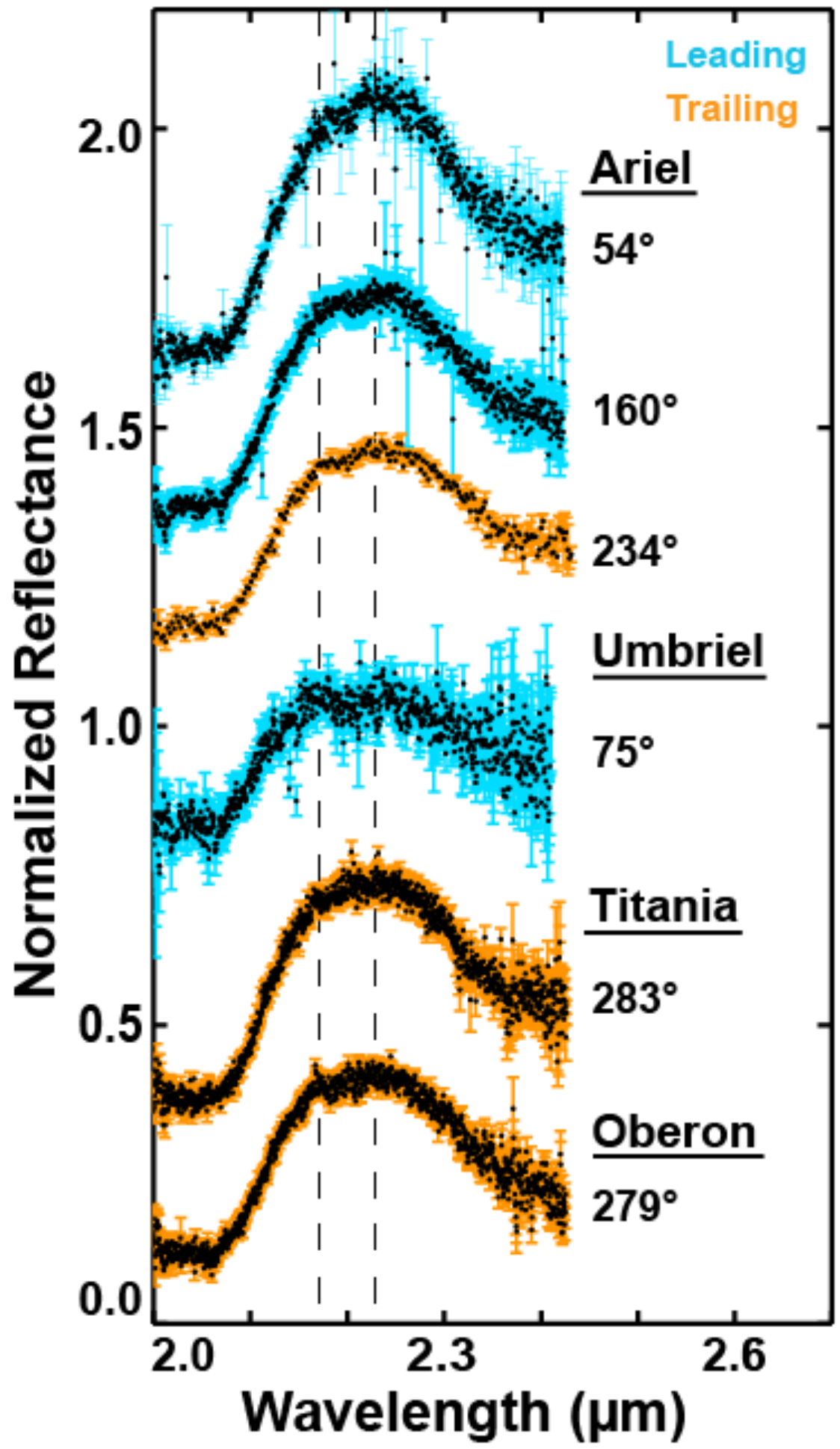

**Figure 5:** SpeX spectra of Ariel, Umbriel, Titania, and Oberon that hint at the possible presence of an $NH_3$-hydrate feature, centered near 2.2 µm, between the two black dashed lines (see Bauer et al. (2002) for a similar feature on Miranda). All spectra gathered in SXD mode, except for Ariel 234°, which was gathered in PRISM mode. Leading (blue error bars) and trailing (orange error bars) hemisphere spectra are indicated.

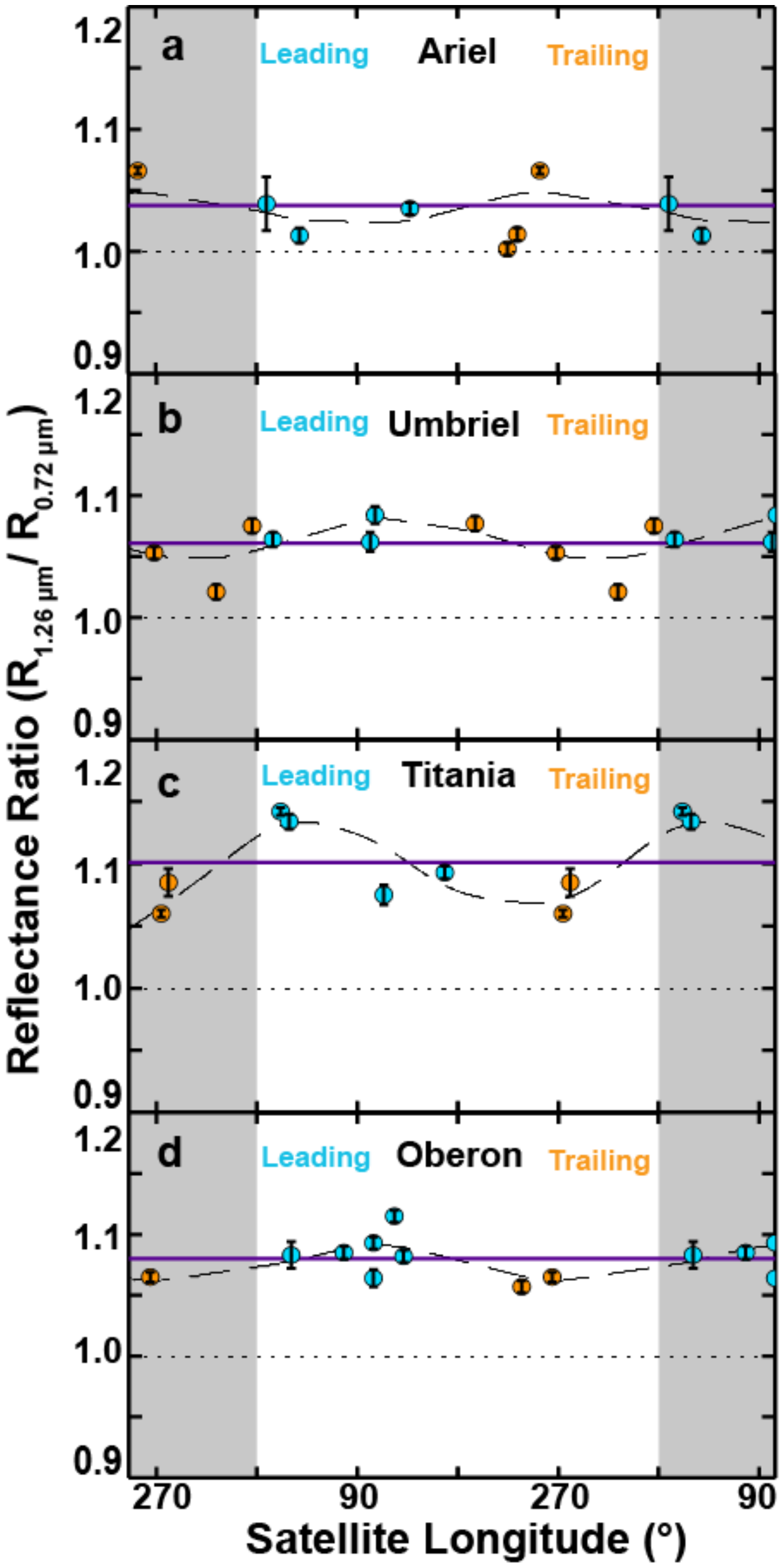

**Figure 6:** Reflectance ratios ($R_{1.26\ \mu m}$/ $R_{0.72\ \mu m}$) as a function of satellite longitude (1$\sigma$ uncertainties) for Ariel, Umbriel, Titania and Oberon (a - d, respectively). Leading (blue) and trailing (orange) hemisphere measurements indicated. Dashed black lines represent sinusoidal fits to the data and solid purple lines show the mean reflectance ratio for each moon. Duplicate longitudes are shown to highlight periodic trends in the abundance of red material on each satellite (gray-toned regions).

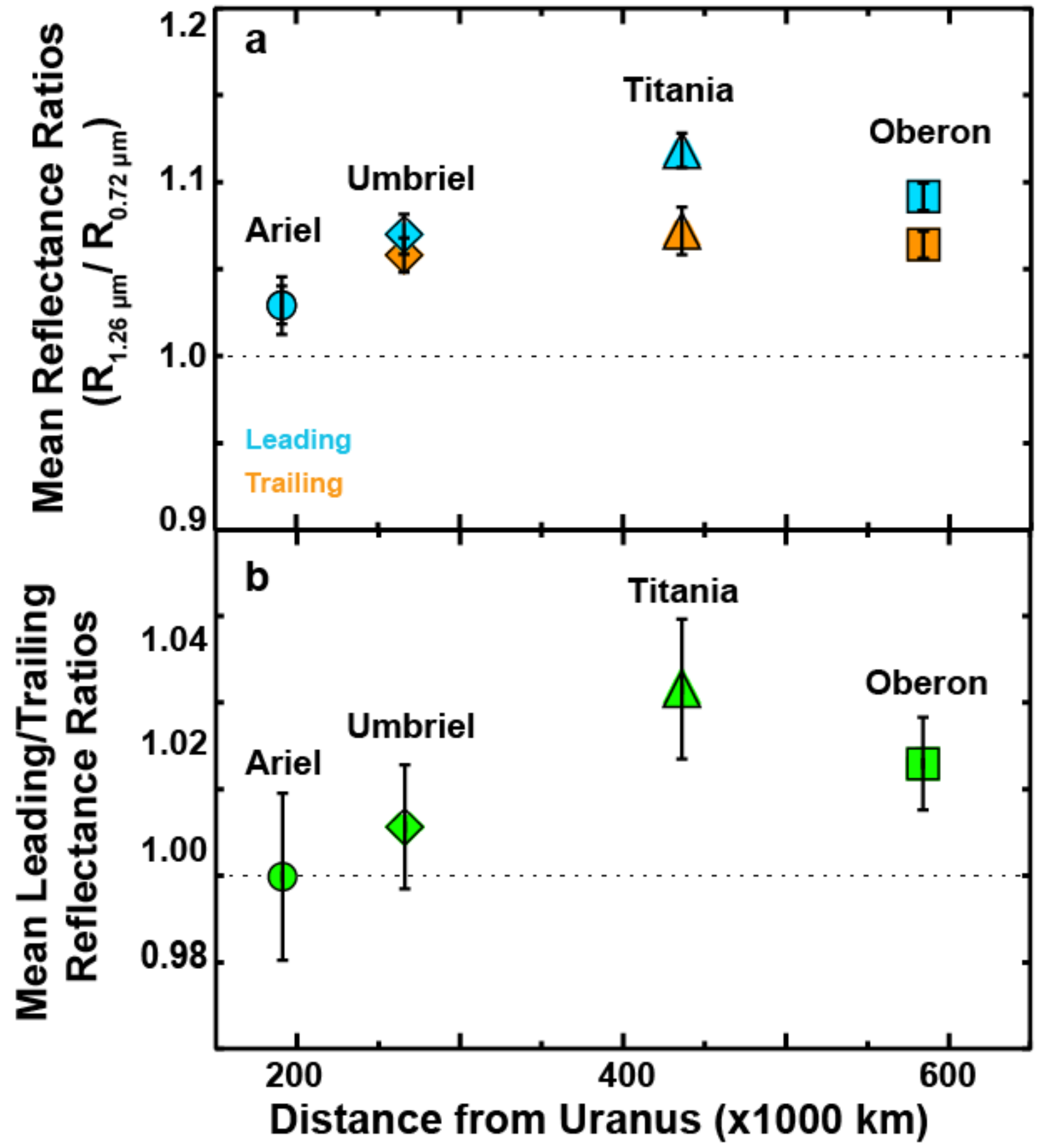

**Figure 7:** (a) Mean reflectance ratios ($R_{1.26\ \mu m}$/ $R_{0.72\ \mu m}$) and $2\sigma$ uncertainties for the leading (blue) and trailing (orange) hemispheres of Ariel (circles), Umbriel (diamonds), Titania (triangles), and Oberon (squares) as a function of planetocentric distance. (b) Mean leading/trailing reflectance ratios for these moons ($2\sigma$ uncertainties), with the same symbology as shown in (a).

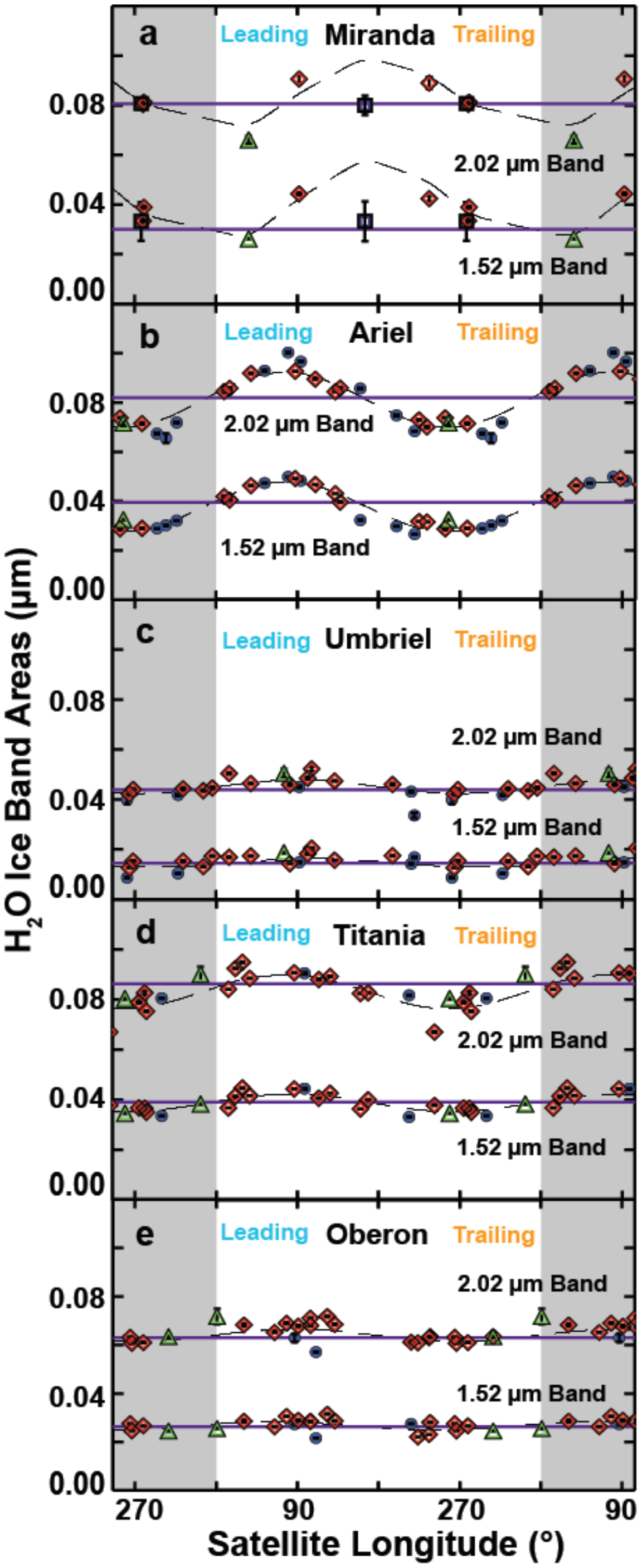

**Figure 8:** Band areas and 1$\sigma$ uncertainties for the 1.52-μm and 2.02-μm $H_2O$ ice bands (bottom and top set of data points in each plot, respectively) as a function of satellite longitude for Miranda Ariel, Umbriel, Titania, Oberon (a - e, respectively). Dashed black lines represent sinusoidal fits to the data and solid purple lines show the mean 1.52-μm and 2.02-μm $H_2O$ band area for each moon. Duplicate longitudes are shown to highlight periodic trends in $H_2O$ ice abundance on each satellite (gray-toned regions). Symbols are the same as shown in Figure 2: Red diamonds (Cartwright), blue circles (Grundy), green triangles (Rivkin), purple squares (Gourgeot).

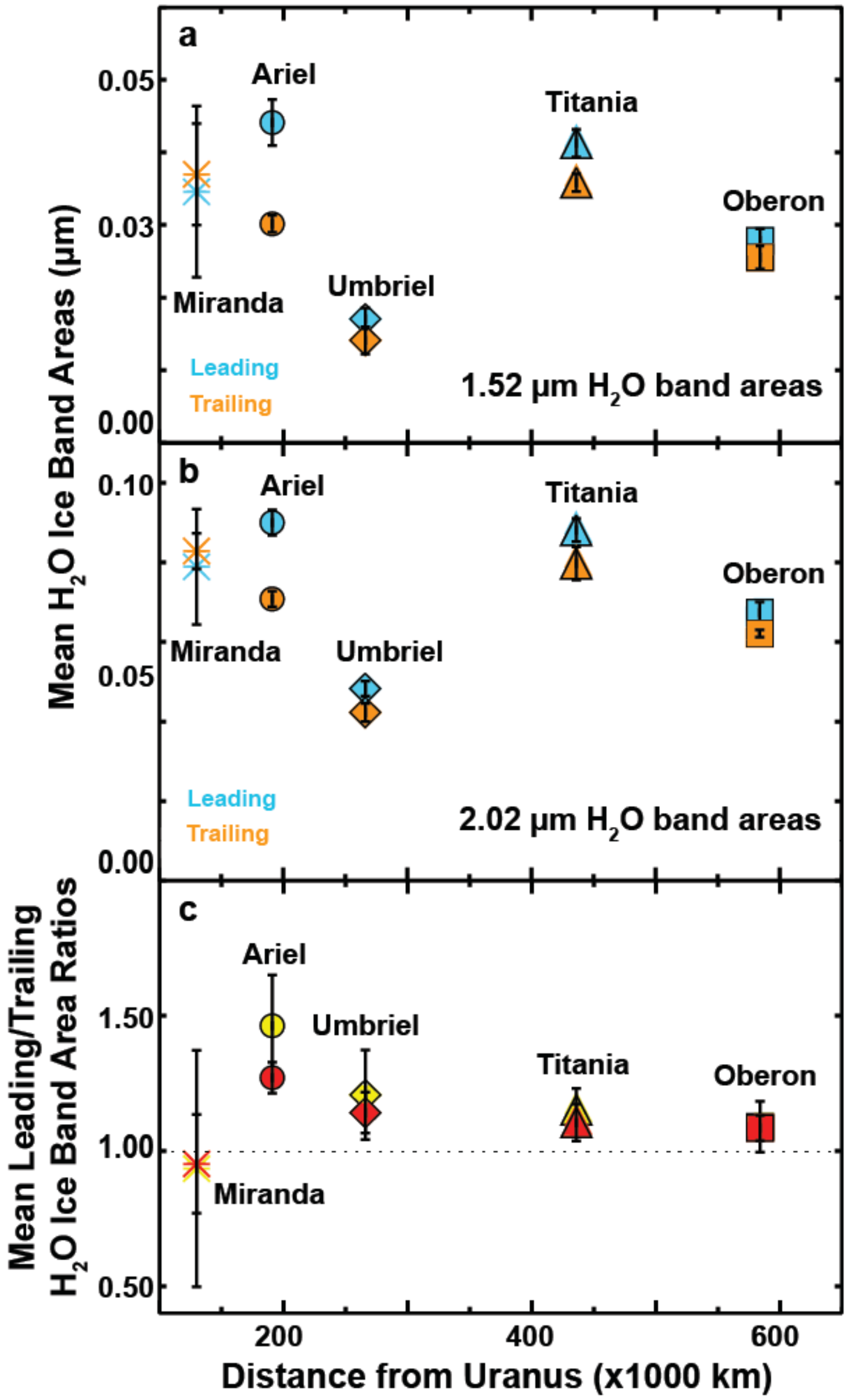

**Figure 9:** Mean band areas and 2$\sigma$ uncertainties for (a) the 1.52-µm and (b) 2.02-µm $H_2O$ ice bands for the leading (blue) and trailing (orange) hemispheres of the Uranian satellites. (c) Mean Leading/Trailing $H_2O$ ice band area ratios for the 1.52-µm (yellow) and 2.02-µm (red) bands (2$\sigma$ uncertainties). In all three plots: Miranda (asterisks), Ariel (circles), Umbriel (diamonds), Titania (triangles), and Oberon (squares).

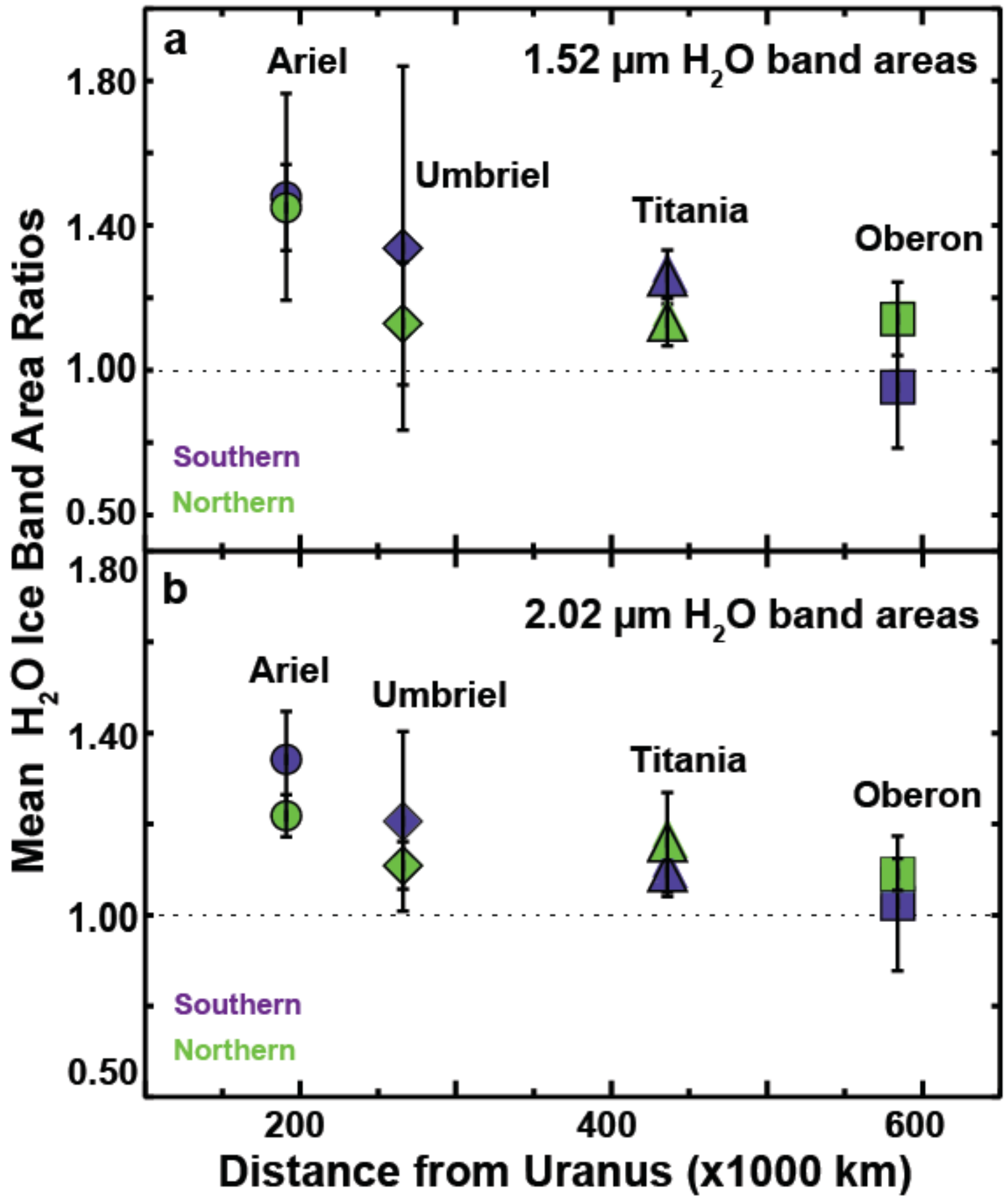

**Figure 10:** Mean leading/trailing $H_2O$ ice band area ratios and $2\sigma$ uncertainties for spectra collected over southern latitudes (purple) and northern latitudes (green) for (a) the 1.52-µm and (b) the 2.02-µm $H_2O$ ice bands. In both plots: Ariel (circles), Umbriel (diamonds), Titania (triangles), and Oberon (squares).

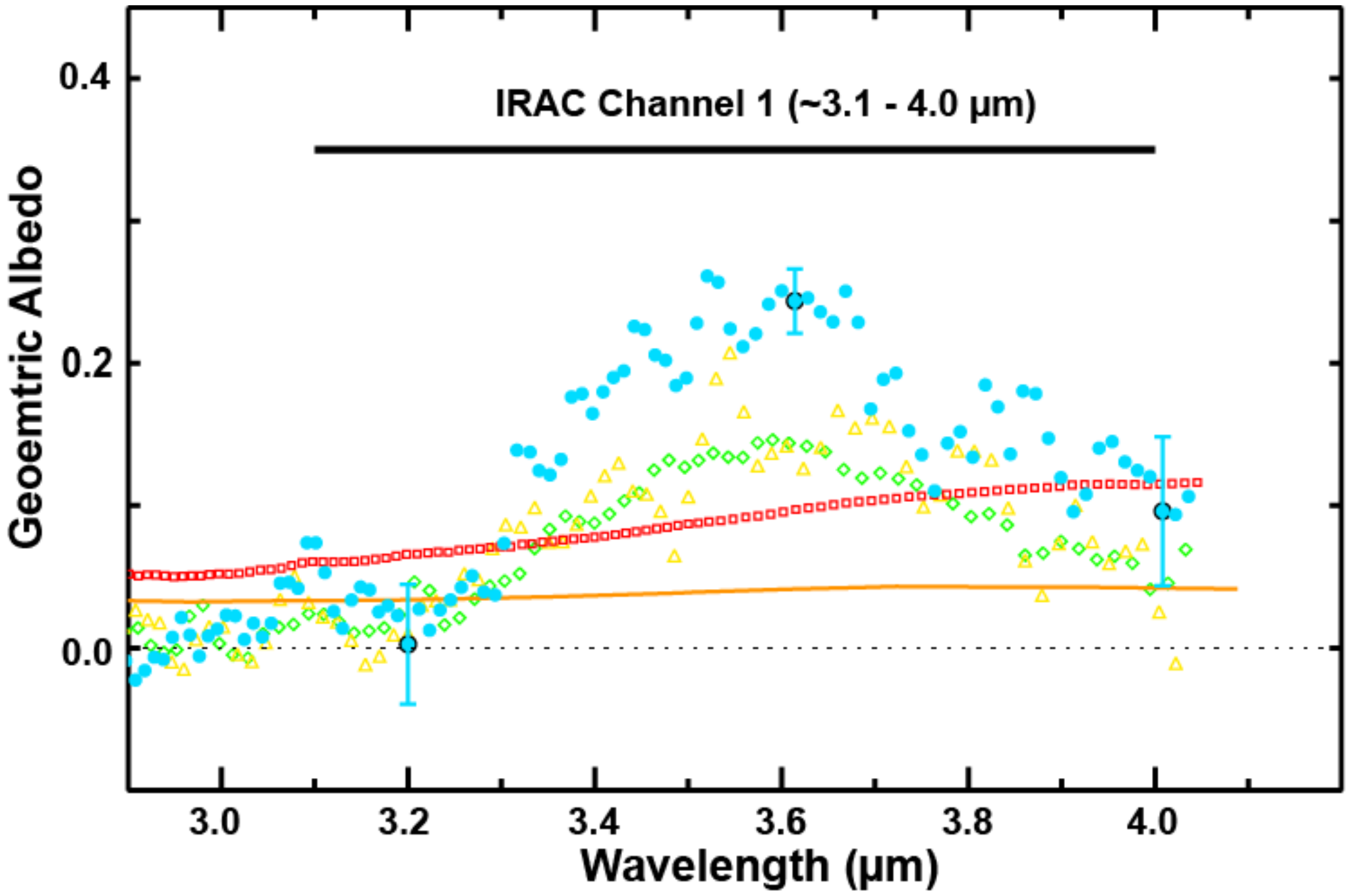

**Figure 11:** Comparison of binned versions of SpeX/LXD spectra of Oberon (light blue circles), Rhea (green diamonds) and Tethys (yellow triangles) (Emery et al., 2005), Callisto (red squares), and a Visual Imaging Mapping Spectrometer spectrum of dark material on Iapetus (solid orange line, published in Pinilla-Alonso et al., 2011). Error bars have been suppressed for clarity. $1\sigma$ errors for the Oberon spectrum at ~3.2, 3.6, and 4.0 μm are shown to illustrate the range in uncertainties ($1\sigma$ uncertainties for the other spectra are comparable or smaller). Wavelength range of IRAC channel 1 is also shown (solid black line).

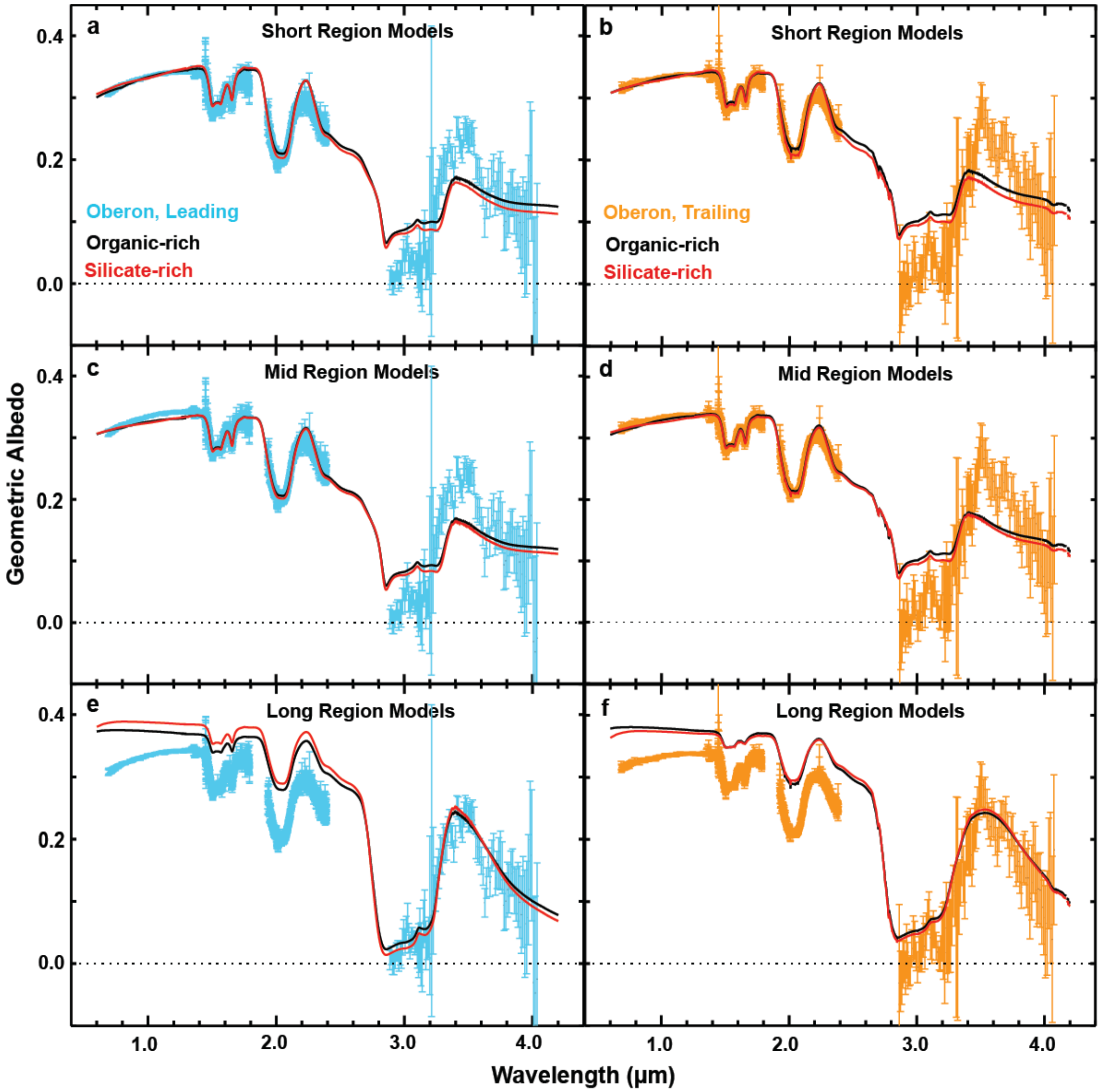

**Figure 12:** Best fit spectral models for the short, mid, and long wavelength regions of Oberon's leading (blue error bars) and trailing (orange error bars) hemisphere. Models including Triton tholins (black) and pyroxene (red) shown for each wavelength region. (see Table 14 for summary of model components and constituent grain sizes).

## Appendix A

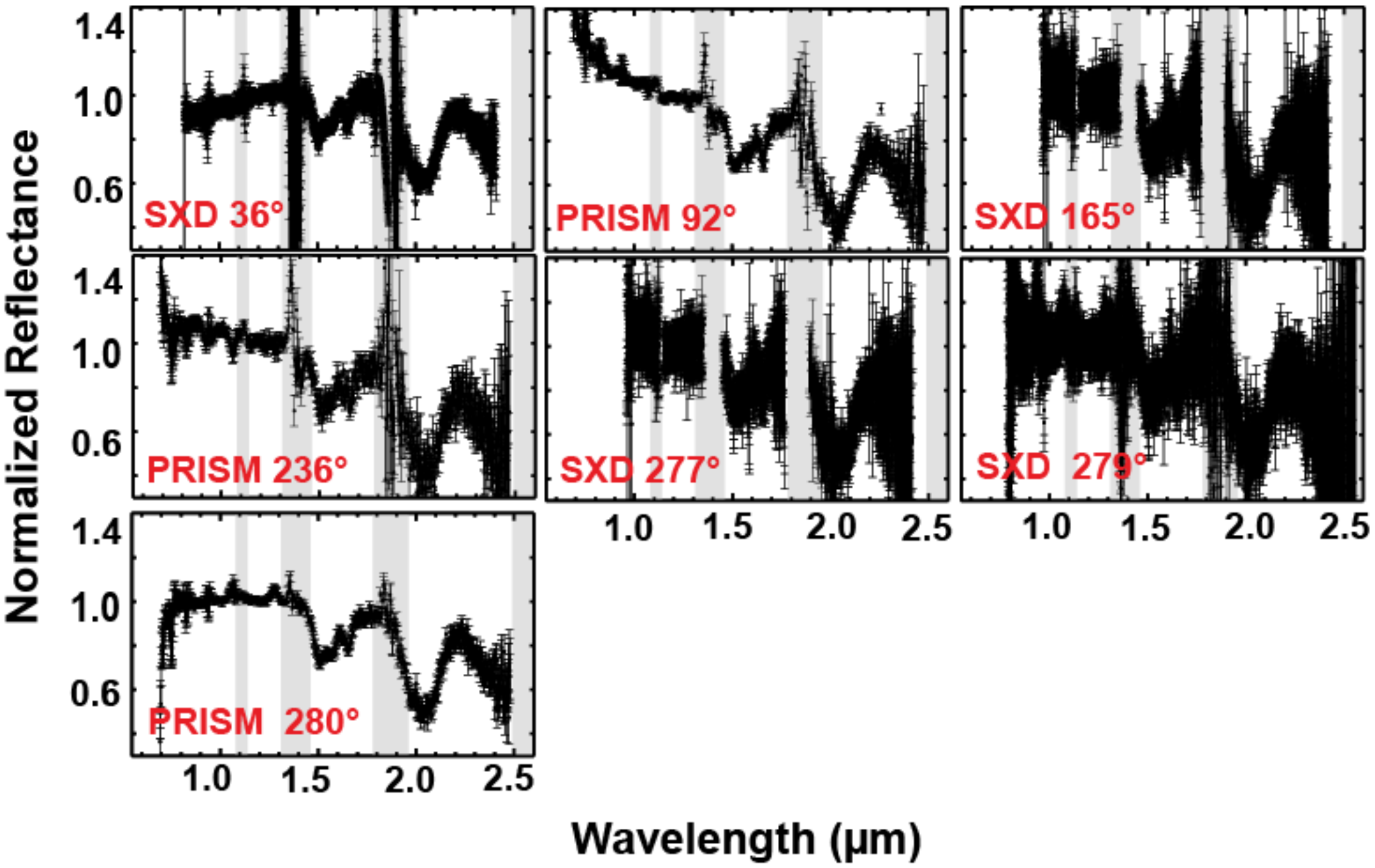

**Figure A1:** Seven SpeX spectra of Miranda (SXD and PRISM modes), organized by increasing mid-observation satellite longitude (listed in bottom lefthand corner of each plot). Each spectrum is normalized to its mean reflectance between 1.20 and 1.22 µm. Wavelength range of telluric bands are indicated by light gray shaded regions.

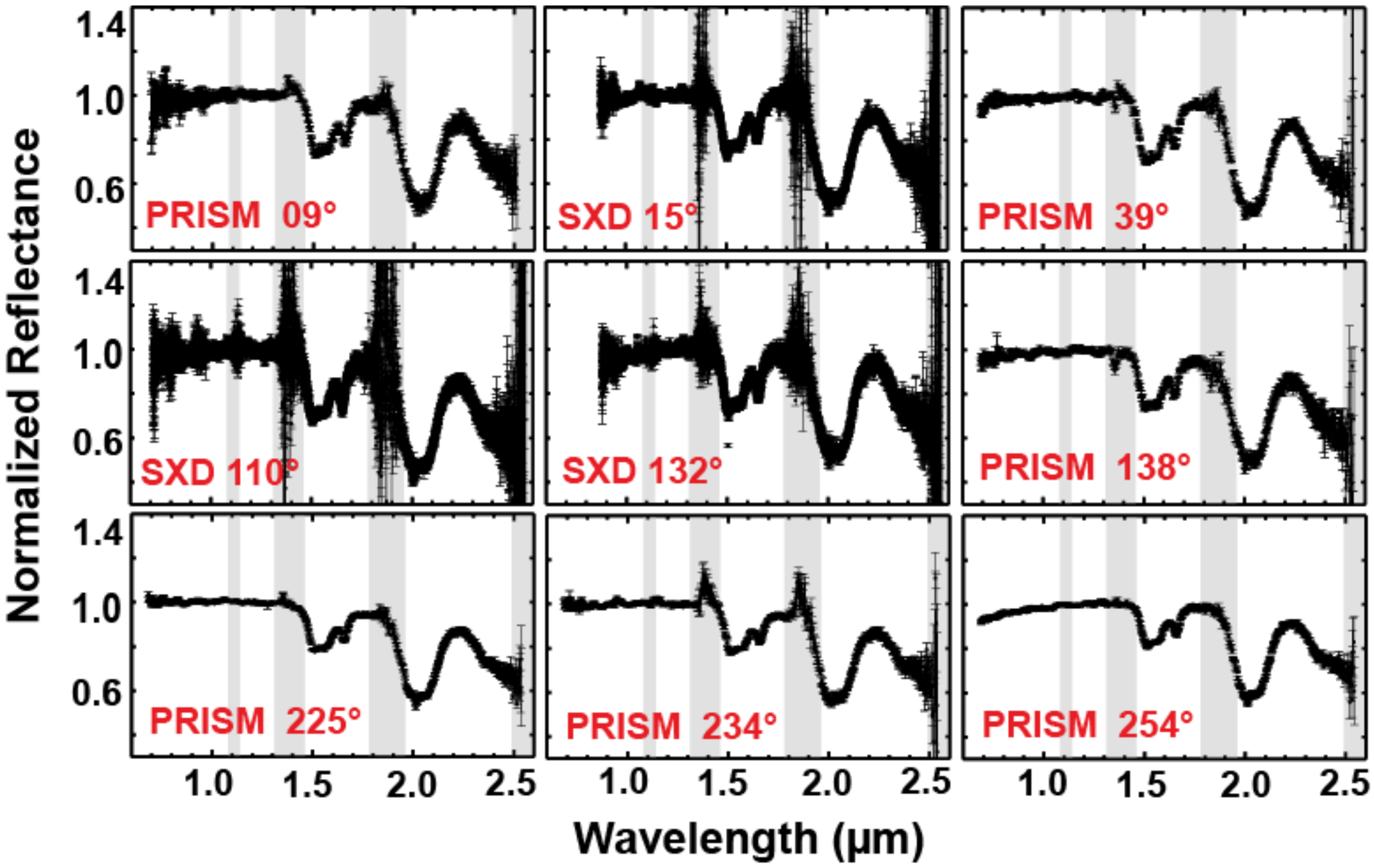

**Figure A2:** Nine SpeX spectra of Ariel (SXD and PRISM modes), organized by increasing mid-observation satellite longitude (listed in bottom lefthand corner of each plot). Each spectrum is normalized to its mean reflectance between 1.20 and 1.22 µm. Wavelength range of telluric bands are indicated by light gray shaded regions.

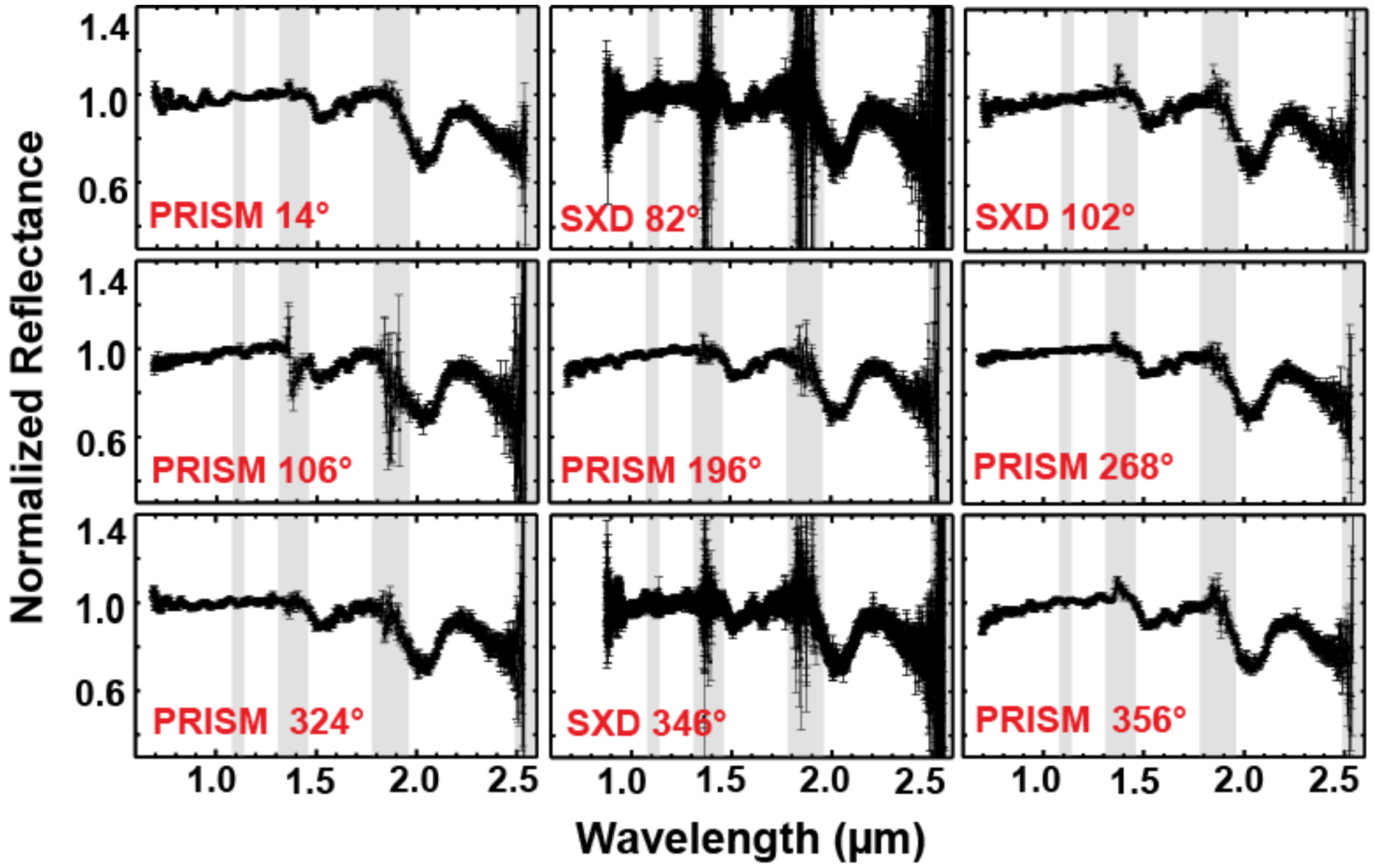

**Figure A3:** Nine SpeX spectra of Umbriel (SXD and PRISM modes), organized by increasing mid-observation satellite longitude (listed in bottom lefthand corner of each plot). Each spectrum is normalized to its mean reflectance between 1.20 and 1.22 μm. Wavelength range of telluric bands are indicated by light gray shaded regions.

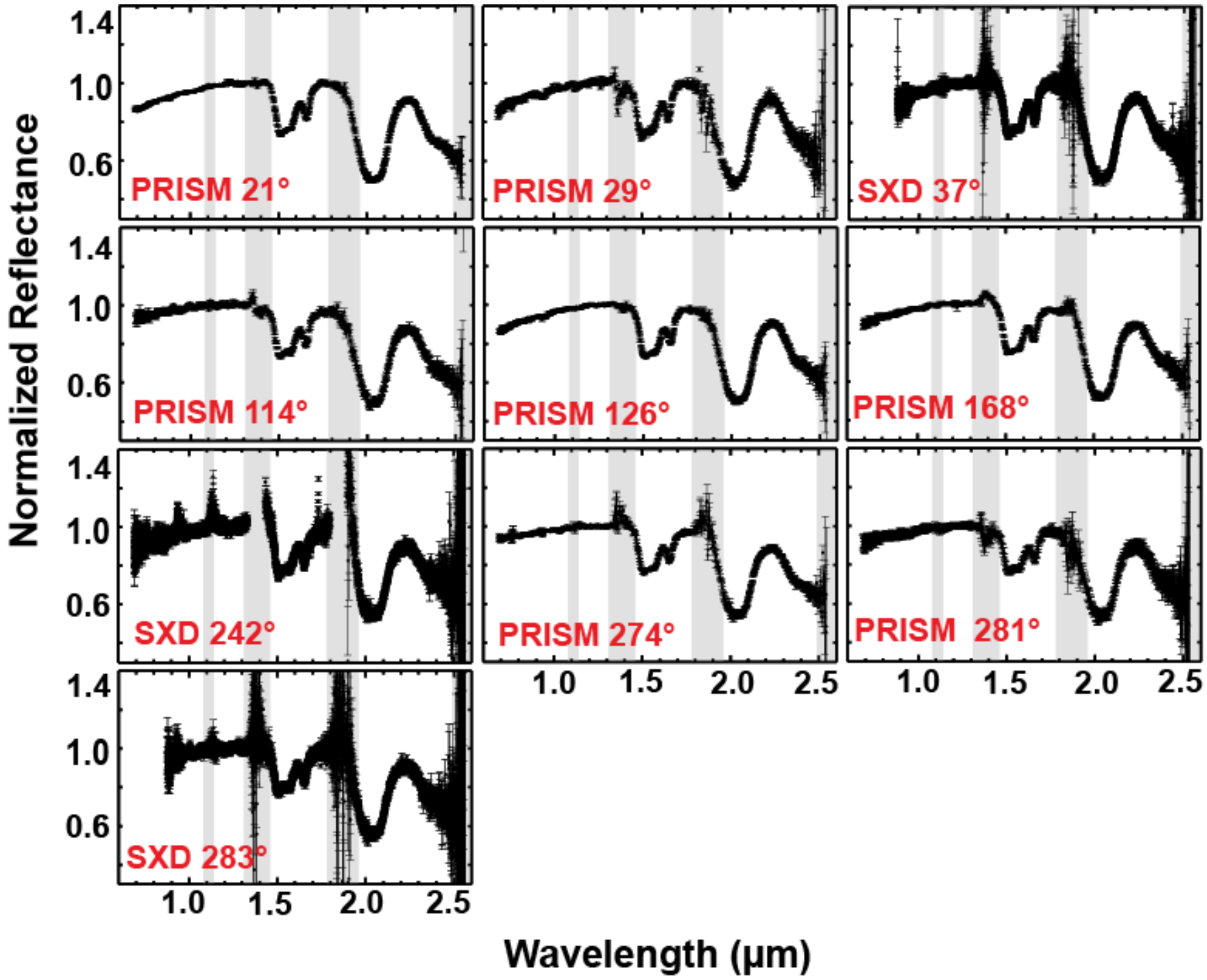

**Figure A4:** Ten SpeX spectra of Titania (SXD and PRISM modes), organized by increasing mid-observation satellite longitude (listed in bottom lefthand corner of each plot). Each spectrum is normalized to its mean reflectance between 1.20 and 1.22 µm. Wavelength range of telluric bands are indicated by light gray shaded regions.

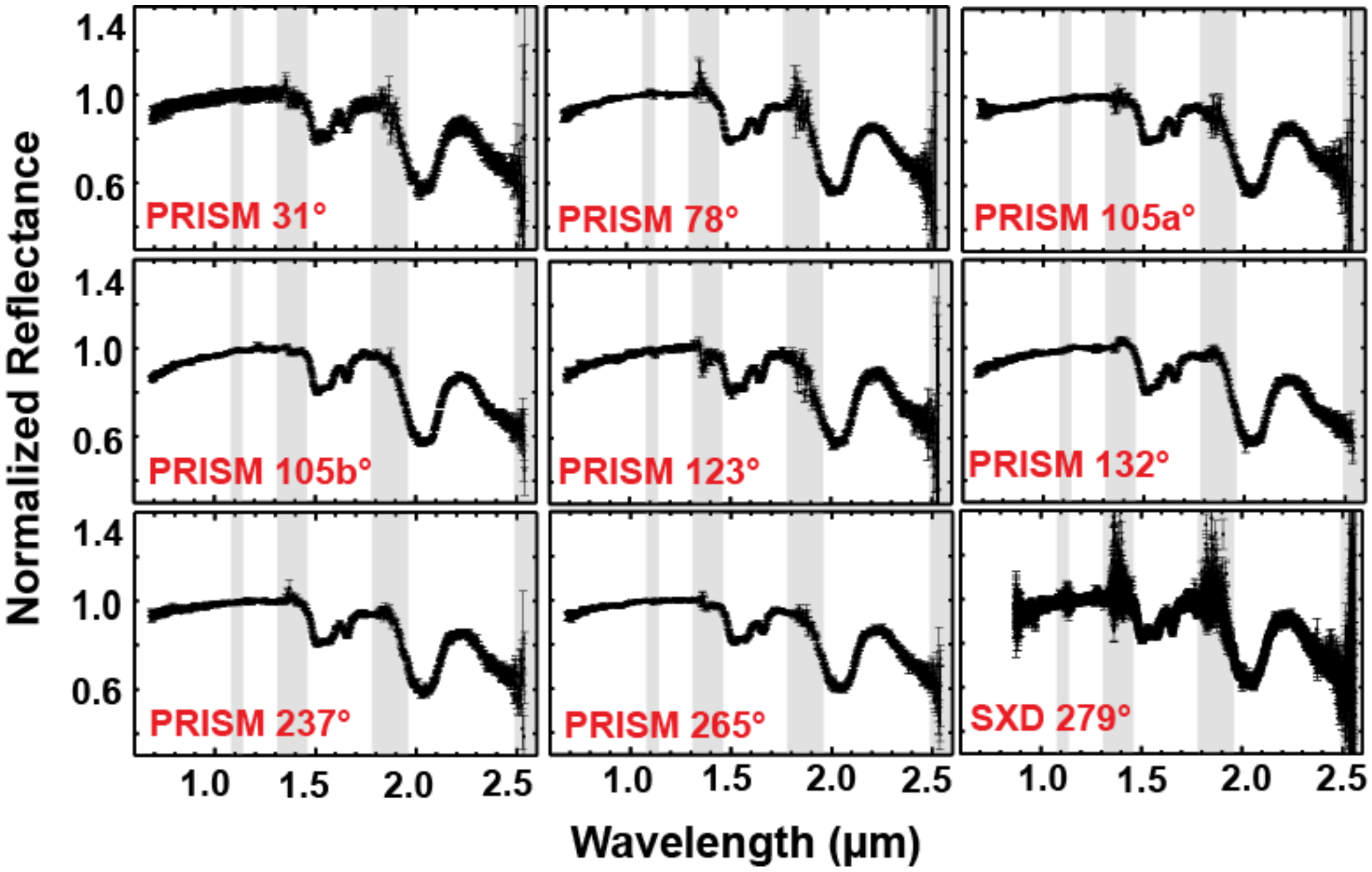

**Figure A5:** Nine SpeX spectra of Oberon (SXD and PRISM modes), organized by increasing mid-observation satellite longitude (listed in bottom lefthand corner of each plot). Each spectrum is normalized to its mean reflectance between 1.20 and 1.22 μm. Wavelength range of telluric bands are indicated by light gray shaded regions.

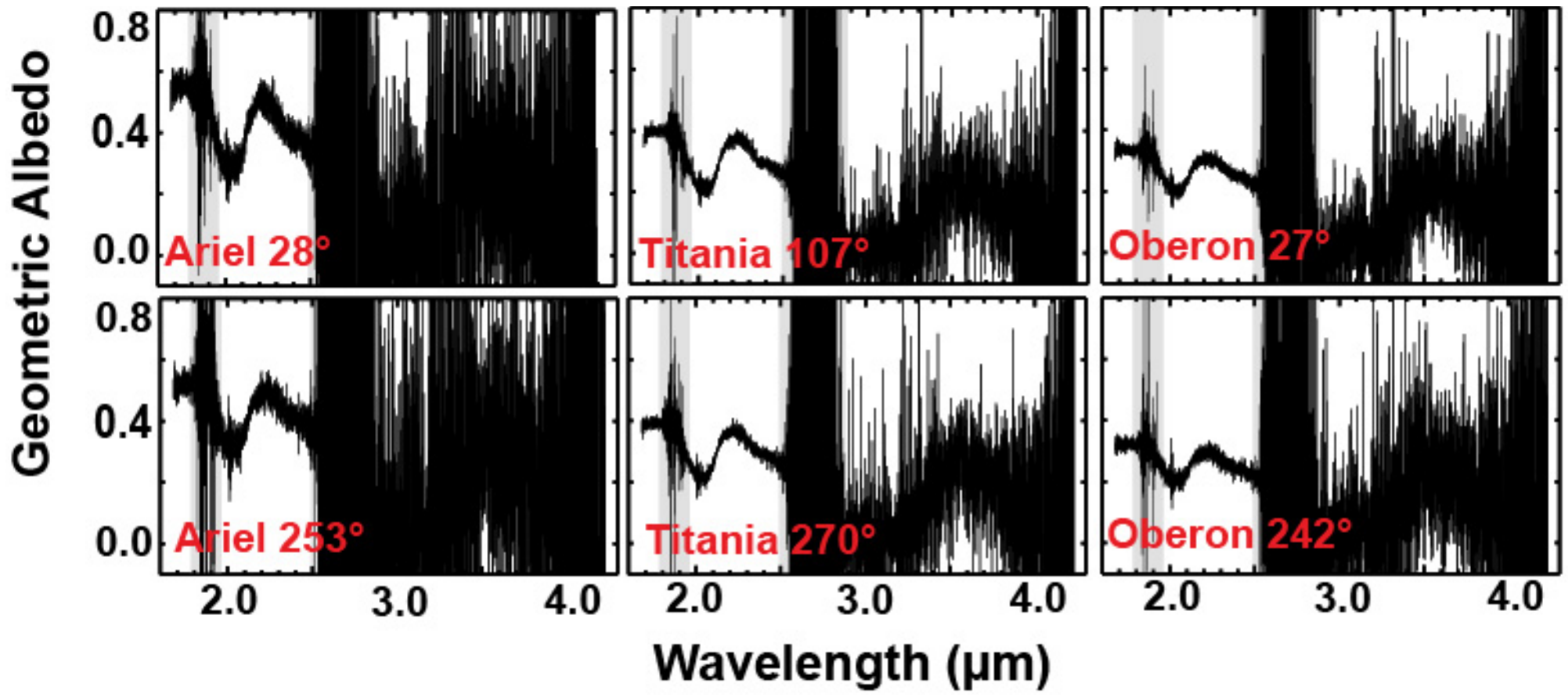

**Figure A6:** Six SpeX spectra (LXD mode) of Ariel, Titania, and Oberon, organized by increasing mid-observation satellite longitude (listed in bottom lefthand corner of each plot). Each spectrum is normalized to its geometric albedo. Wavelength range of telluric bands are indicated by light gray shaded regions.